\documentclass[fleqn,usenatbib]{mnras}

\usepackage{newtxtext,newtxmath}
\usepackage{xcolor}

\usepackage[T1]{fontenc}

\DeclareRobustCommand{\VAN}[3]{#2}
\let\VANthebibliography\thebibliography
\def\thebibliography{\DeclareRobustCommand{\VAN}[3]{##3}\VANthebibliography}

\usepackage{graphicx}	
\usepackage{amsmath}	
\usepackage[cal=boondoxo]{mathalfa}

\title[{\tt XSPEC} models of reflection near compact objects]{Simple numerical X-ray polarization models of reflecting axially symmetric structures around accreting compact objects}

\author[J. Podgorn\'y et al.]{
J. Podgorn\'y$^{1,2,3}$\thanks{E-mail: jakub.podgorny@asu.cas.cz},
M. Dov{\v{c}}iak$^{2}$ 
and F. Marin$^{1}$ 
\\
$^{1}$Universit\'e de Strasbourg, CNRS, Observatoire Astronomique de Strasbourg, UMR 7550, F-67000 Strasbourg, France\\
$^{2}$Astronomical Institute, Academy of Sciences of the Czech Republic, Bo{\v{c}}n\'i II, CZ-14131 Prague, Czech Republic\\
$^{3}$Astronomical Institute, Charles University, V Hole{\v{s}}ovi{\v{c}}k\'ach 2, CZ-18000 Prague, Czech Republic\\
}

\date{Accepted XXX. Received YYY; in original form ZZZ}

\pubyear{2023}

\begin{document}
\label{firstpage}
\pagerange{\pageref{firstpage}--\pageref{lastpage}}
\maketitle

\begin{abstract}
We present a series of numerical models suitable for X-ray polarimetry of accreting systems. Firstly, we provide a spectropolarimetric routine that integrates reflection from inner optically thick walls of a geometrical torus of arbitrary size viewed under general inclination. In the studied example, the equatorial torus is illuminated by a central isotropic source of X-ray power-law emission, representing a hot corona. Nearly neutral reprocessing inside the walls is precomputed by Monte Carlo code {\tt STOKES} that incorporates both line and continuum processes, including multiple scatterings and absorption. We created a new {\tt XSPEC} model, called {\tt xsstokes}, which in this version enables efficient X-ray polarimetric fitting of the torus parameters, observer's inclination and primary emission properties, interpolating for arbitrary state of primary polarization. Comparison of the results to a Monte Carlo simulation allowing partial transparency shows that the no-transparency condition may induce different polarization by tens of \%. Allowing partial transparency leads to lower/higher polarization fraction, if the resulting polarization orientation is perpendicular/parallel to the rotation axis. We provide another version of {\tt xsstokes} that is suitable for approximating nearly neutral reflection from a distant optically thick disc of small geometrical thickness. It assumes local illumination averaged for a selected range of incident angles, representing a toy model of a diffuse corona of various physical extent. Assessing both {\tt xsstokes} variants, we conclude that the resulting polarization can be tens of \% and perpendicularly/parallelly oriented towards the rotation axis, if the reflecting medium is rather vertically/equatorially distributed with respect to a compact central source.
\end{abstract}




\section{Introduction}\label{introduction}

X-ray polarimetry is undergoing a rebirth due to the successful launch and operation of the \textit{Imaging X-ray Polarimetry Explorer} (\textit{IXPE}) \citep{Weisskopf2022}. Out of many scientific cases of X-ray polarimetry, where reflection from axially symmetric structures plays a role, we name the extragalactic active galactic nuclei (AGN) with a central supermassive black hole, and the Galactic accreting stellar-mass black holes, neutron stars and white dwarfs. Apart from the polar scatterers, such as jets and ionization cones, the compact accreting objects are often surrounded by distant equatorial reprocessing material that changes the properties of the central X-ray emission. This can be the broad line regions or the dusty parsec-scale torus of AGNs \citep{Antonucci1993,Urry1995}, or equatorial outflows arising from outer accretion disc surrounding the supermassive black holes or orders-of-magnitude lighter objects \citep[see e.g.][]{Koljonen2020,Neilsen2020,Miller2020, Ratheesh2021}. A hot coronal plasma producing primary X-ray power-law emission is located in the inner-most accretion regions \citep{Sunyaev1980, Haardt1991, Haardt1993}. Various geometries of the corona are continuously considered in literature with theoretical polarization typically up to a few \% \citep{Beheshtipour2017, Tamborra2018, Marinucci2018, Poutanen2018, Krawczynski2022b, Ursini2022}. These predictions are observationally confirmed by \textit{IXPE} with a preference for equatorially elongated coronae for particular sources \citep{Krawczynski2022, Marinucci2022, Tagliacozzo2023, Ingram2023, Gianolli2023, Veledina2023b}.

The spectropolarimetric modelling of distant reprocessing of the coronal emission reaches a range in complexity and computational effectivity. From analytical and semi-analytical models \citep[e.g.][]{Chandrasekhar1960, Brown1977, Sunyaev1985, Matt1993b, Poutanen1996, Poutanen1996b, Bianchi2010, Veledina2023} to complex radiative transfer equation solvers \citep{ Matt1993,Dorodnitsyn2010, Dorodnitsyn2011} to Monte Carlo (MC) simulations \citep[e.g.][]{Matt1989, Ghisellini1994, Matt1996, Goosmann2011, Marin2016b,Marin2018b, Marin2018c, Ratheesh2021, Ursini2023, Tomaru2024}. Admittedly, the new models presented in this paper are serving as first-order predictions and represent a largely simplistic, yet computationally efficient scenario. No special- or general-relativistic effects are taken into account. We study a nearly neutral toroidal geometry that reflects a point-like isotropic power-law emission located in the center. Only a single reprocessing at one part of the optically thick inner walls is taken into account, although locally, the effects of multiple scatterings and energy-dependent absorption are treated. Across the geometrical torus surface, we numerically integrate precomputed local spectropolarimetric tables that were originally developed for reflection from accretion discs in AGNs. We refer the reader to \cite{Podgorny2021} for a full description of these tables. The aim of this work is to achieve an approximative model enabling fast estimates with the data fitting tool {\tt XSPEC} \citep{Arnaud1996}. We compare these results to more sophisticated MC results of partially ionized equatorial obscurers presented in \cite{Podgorny2023b} (hereafter Paper I) that used the {\tt STOKES} code \citep{Goosmann2007,Marin2012,Marin2015,Marin2018} for the same radiative transfer problem self-consistently, i.e. without the need for precomputation and assuming some level of transparency of the obscurer. In this way, we can show that the allowed relative contribution of reflection on the inner walls of an accreting funnel determines the output polarization properties to a large extent, depending on geometrical properties of the scattering region and primary radiation characteristics.

Apart from a general numerical integrator across a toroidal structure, we use a convertor of American Standard Code for Information Interchange (ASCII) tables to Flexible Image Transport System (FITS) format \citep{FITS} in the Office for Guest Investigator Programs (OGIP) standard \citep{OGIP} suitable for {\tt XSPEC} in its latest version that includes polarization. This results in the above mentioned torus reflection model suitable for {\tt XSPEC}.

One other variant of the {\tt XSPEC} model is presented, achieved in a similar manner. It approximates net polarization from centrally illuminated faraway regions of cold matter distributed near the equatorial plane with low height relative to the radius. Given the averaging across an arbitrary range of impinging angles of the local reflection tables computed in \cite{Podgorny2021}, this model approximates the emitter of X-ray power-law, situated inside and/or above the disc plane with low to high extension in physical size.

The paper is organized as follows. In Section \ref{methods} and Appendices \ref{technical_details}, \ref{torus_calculus} and \ref{technical_disc} we describe the new models developed, i.e. the torus reflection integrator, the table conversion routine, and the two {\tt XSPEC} fitting models: for reflection from a torus and for reflection from a distant disc. In Section \ref{compare_xsstokes} we discuss the results of the torus reflection model and compare them to a detailed MC approach and to an analytical approach. In Section \ref{distant_disc} we discuss the results of the disc reflection model and compare them to an analytical approach. The dependency of the resulting polarization state on incident polarization state for both studied geometries is further examined in Appendix \ref{incident_polarization}. In Section \ref{conclusions} we conclude.

\section{Modelling}\label{methods}

We define the linear polarization degree $p$ and polarization position angle $\Psi$ in the usual way from the Stokes parameters $I$, $Q$ and $U$
\begin{equation}\label{ppsidef}
	\begin{aligned}
		p &= \dfrac{\sqrt{Q^2+U^2}}{I} \\
		\Psi &= \dfrac{1}{2}\textrm{\space}\arctan_2\left(\dfrac{U}{Q}\right)   \textrm{ ,}
	\end{aligned}
\end{equation}
where $\arctan_2$ denotes the quadrant-preserving inverse of a tangent function and $\Psi = 0$ means that the polarization vector is oriented \textit{parallel} to the system axis of symmetry projected to the polarization plane. $\Psi$ increases in the counter-clockwise direction from the point of view of an incoming photon. Due to symmetry, the polarization genesis from axially symmetric reflecting surface is such that the observed photons are polarized on average parallel or perpendicular to the main axis. Thus, we will use the notation of positive or negative polarization degree $p$ for such resulting $\Psi$, respectively. When describing the polarization of the source emission, we will use $\Psi_0$ and $p_0$ -- with the same positive or negative notation for polarization fraction inside figures, as only parallelly or perpendicularly oriented polarization with respect to the principal axis will be used in the provided examples. Note that due to general-relativistic effects in the close vicinity of the compact object, this might not always be the case for realistic central emission \citep[see e.g.][]{Krawczynski2022b, Ursini2022}.

\subsection{Simple toroidal reflector}\label{torus_reflection}

Paper I introduced various geometrical archetypes for studying reprocessing of with a power-law index $\Gamma$ in a distant equatorial scattering region. The simulations were done using the {\tt STOKES} code: an MC method with various X-ray line and continuum processes implemented with emphasis on polarization computations. The code takes into account all major fluorescent and resonant lines in the X-ray band. It takes into account photo-electric absorption, as well as multiple Compton down-scatterings. Compton up-scatterings are neglected in the used version. A broad parametric range was explored in Paper I, including the effects of partial transparency and partial ionization. Here we will take just part of this work and compare it to a much simpler approach, trying to capture the main polarization driving mechanisms. We will use an identical geometrical setup to the Case C described in Paper I for easy one-to-one comparisons. This means a perfectly circular torus profile ($a = b$ subcase of an elliptical torus, see Paper I for the model parameters and details) with a half-opening angle $\Theta$ viewed under some inclination $i$ from the pole. Such obscurer has a convex inner boundary with respect to the central source. The emission is isotropic, and is located in the center of the torus and is approximated as a point source. Note that this is again a huge simplification to more realistic hot X-ray emitting coronae near the central accreting engine (see references in Section \ref{introduction}). Direct emission, i.e. light that arrives to the observer without any interactions, is not added to the reflected component.

\begin{figure}
	\includegraphics[width=1.\columnwidth]{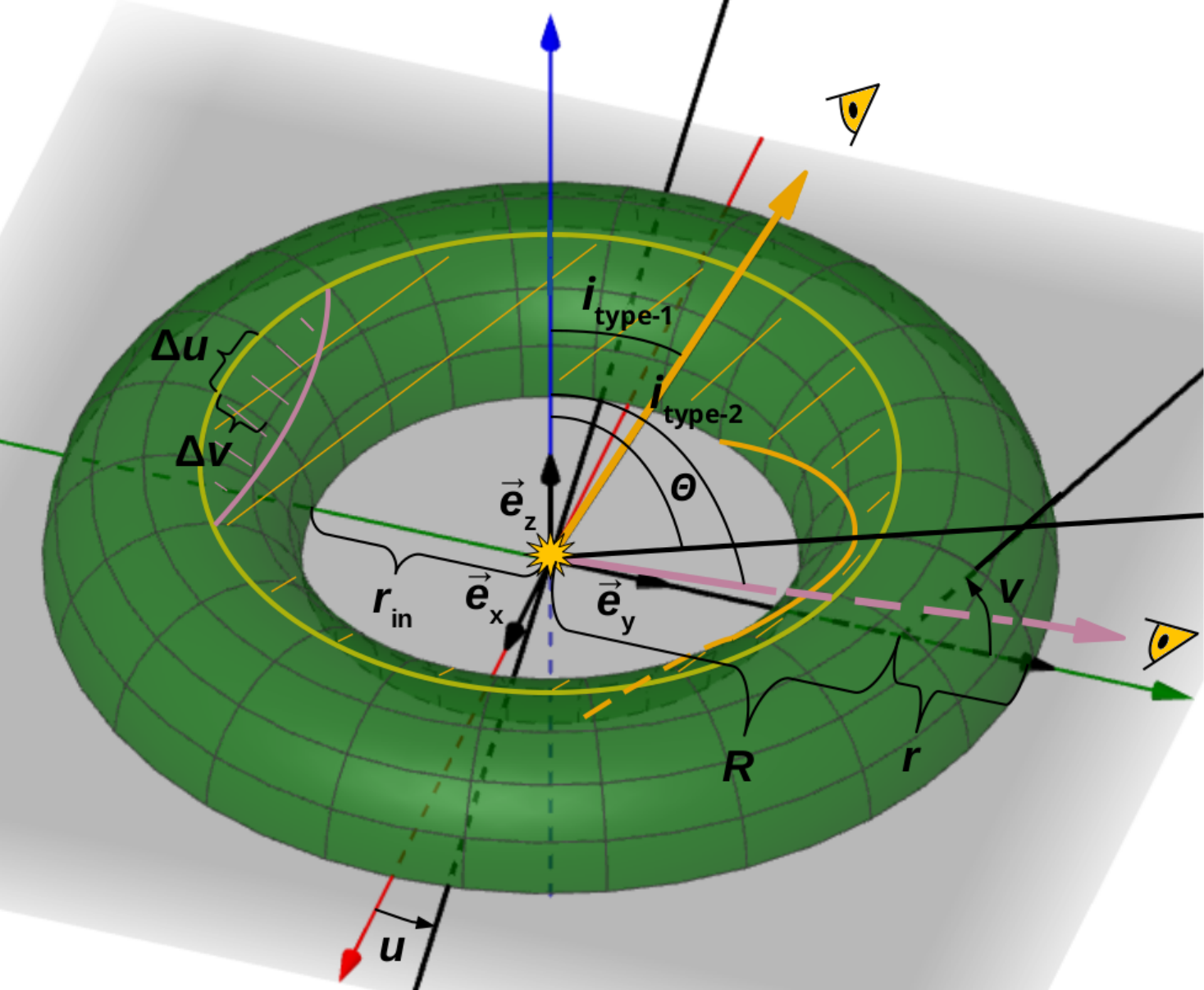}
	\caption{A circular toroidal reflector that is considered for the results presented in Section \ref{compare_xsstokes}. Being located in the center of the coordinate system, the X-ray power-law of arbitrary incident polarization and power-law index $\Gamma$ is isotropically illuminating the inner side of the torus surface. Here we show only the upper half-space above the equatorial plane for clarity, but the model allows to choose if only the upper part of the torus above the equator (with $z>0$) may be illuminated, or if the surface below the equator (with $z <0$) is also taken into account. The observer is viewing the axially symmetric system in the $yz$-plane and may be generally inclined under inclination $0\degr < i < 90\degr$ measured from the rotation axis. The material reflects once and we only account for the surface regions that are directly illuminated by the central source, i.e. here in the upper half-plane represented by the yellow shaded area. The illuminated part has to be in addition visible in the observer's line of sight to be accounted for in the integration. The lower boundaries, i.e. the terminators, of the areas that are obscured by the opaque torus for an observer in two generic $i_\textrm{type-1}$ and $i_\textrm{type-2}$ inclinations are schematically drawn in orange and pink colors, respectively. The visible reflecting areas for a generic type-1 and type-2 observer are then shaded in orange and pink, respectively. The torus parameters $r$, $R$ and $r_\textrm{in}$ and the $u$ and $v$ surface coordinates are in detail described in Appendix \ref{torus_calculus}.}
	\label{xsstokes_mo}
\end{figure}

Figure \ref{xsstokes_mo} shows the torus reflection setup. For clarity, we show only the upper half-plane of the torus, even though all the results shown in this paper take the reflecting area below the equator into account as well, if not mentioned otherwise. We refer to Appendices \ref{technical_details} and \ref{torus_calculus} for detailed numerical implementation of the resulting {\tt XSPEC} model {\tt xsstokes}, in this variant named {\tt xsstokes\_torus}, and description of the corresponding routines. The assumption of axial symmetry allows swift interpolation for arbitrary state of the primary polarization given by $p_0$ and $\Psi_0$. Only a single reprocessing from the surface is calculated. In other words we only account for those photons that reach the inner walls of the torus from the central source, they are locally reprocessed by means of precomputed nearly neutral multiple-scattering reflection tables, they emerge at the same surface location, and reach the observer under the assumed global inclination angle $i$. No self-irradiation of the torus is allowed, nor partial transparency effects. The torus is optically thick. This allows to study the impact of multiple scatterings between different torus surface regions when the photon escapes the scattering medium in between two scattering events and the impact of partial transparency of the surface layers near grazing angles that were both accounted for in Paper I.

The central point-like source illuminates up to the tangent point between the torus and a photon direction emitted in a Newtonian picture from the center under the half-opening angle $\Theta$, i.e. the shadow boundary given by the opaque toroidal structure. The viewing angle given by $i$ can span any value from type-1 ($i < \Theta$) to type-2 ($i > \Theta$) values. But note that the visible fraction of the illuminated part of the inner surface gets shrinked considerably for high inclinations.\footnote{For a sketch see the orange and pink lines in Figure \ref{xsstokes_mo}, representing the surface terminators, i.e. the boundaries of the visible surface of the inner part of the torus for a particular type-1 and type-2 viewing angle. The reflecting region that the integrating routine accounts for to compute the total output is then given by the intersection of the illuminated part (given by the torus half-opening angle grazing boundary -- displayed in yellow in Figure \ref{xsstokes_mo}) and the visible part to an observer inclined at $i$.} This allows to estimate the polarized flux contribution from reflection off the opposite side of the torus for highly obscured type-2 AGNs or ultra-luminous X-ray sources (ULX) \citep{Revnivtsev2002, Fabrika2004, Abramowicz2005, Casares2014, Kaaret2017, Motta2017b, Motta2017a, Jiang2019}.

For the local reflection at each infinitesimal surface area (in the code numerically approximated by a small rectangular area with an orientation locally tangent to the torus surface) on the inner illuminated part of the torus surface, we used the most neutral version of the X-ray spectropolarimetric reflection tables created for constant-density plane-parallel accretion discs of AGNs. These tables, fully described in \cite{Podgorny2021}, were computed combining the non-LTE radiative transfer iterator {\tt TITAN} \citep{Dumont2003} and {\tt STOKES}. The reflection includes multiple Compton down-scatterings and photoelectric absorption with high accuracy. The X-ray illuminated partially ionized slab is computed all the way until optical depths of $\tau_\textrm{e} \approx 7$, while we only count the photons that are eventually not absorbed and escape back to the surface. Hence, we obtain a more detailed ``reflection'' (a reprocessing) from an infinitesimal surface on the torus than by a single-scattering reflector.

\subsection{Simple disc reflector}\label{disc_reflection}

The second version of the {\tt xsstokes} model, named {\tt xsstokes\_disc}, is intended to approximate X-ray polarization due to reflection from a static nearly neutral disc in the equatorial plane for unobscured AGNs. Its numerical implementation, which follows similar steps to the torus reflection variant, is described in Appendix \ref{technical_disc}. Figure \ref{xsstokes_disc_sketch} shows the examined reflection geometry. We assume uniform integration (i.e. with no weighting) of the same nearly neutral local reflection tables from \cite{Podgorny2021} in all azimuthal emission angles, $\Phi_\textrm{e}$, and in a selected range of high incident angles, $\delta_\textrm{i}$: $0 \leq \mu_\textrm{i} = \cos{\delta_\textrm{i}} \leq M_\mathrm{i}$, where $M_\mathrm{i} \in [0.2;1]$. Since the incident angle $\delta_\textrm{i}$ is measured from the disc normal, the upper limit $M_\mathrm{i}$ on its cosine acts as a scaling factor on the coronal physical size. If $M_\mathrm{i} = 1$, we integrate over the entire range of impinging angles, meaning the corona is significantly extended, reaching above the disc possibly also in the radial direction from the central compact object. The choice of low $M_\mathrm{i}$ ($\lesssim 0.4$) represents a case of a radially distant or truncated disc, as if it was illuminated by e.g. a compact central corona or hot inner accretion flow with low or moderate vertical extension located in the central parts, and/or if the disc in the outer rings had a considerable changing elevation with respect to the equatorial plane. But note that the latter scenario, if the reflecting medium deviates non-linearly from the equatorial plane, produces a range of emission inclination angles for a particular distant observer. Because we keep a particular emission angle $i$ (adopted as $\delta_\textrm{e}$ from the original local reflection tables) as a free parameter, which drives the polarization results to a significant extent, such thick disc scenarios should be rather approximated by the reflecting torus model described in the previous section. The remaining free parameters of the reflecting disc model are then the power-law index $\Gamma$ and arbitrary incident polarization $p_0$ and $\Psi_0$, for which we interpolate in the same way as for the torus reflection.
\begin{figure}
	\includegraphics[width=\columnwidth]{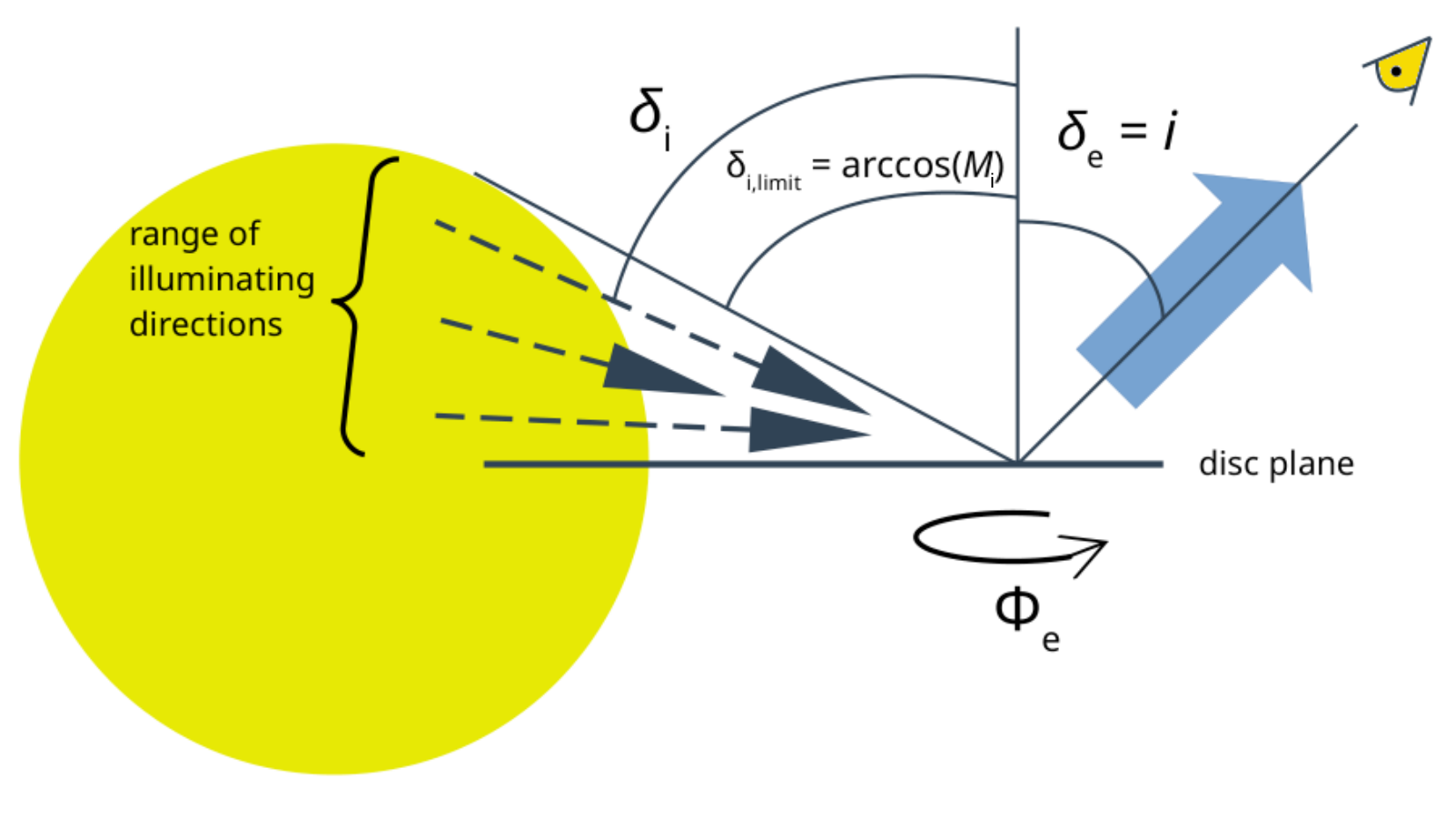}
        \centering
	\caption{The reflection geometry considered for the results presented in Section \ref{distant_disc}. A distant nearly neutral geometrically thin disc residing in the equatorial plane is illuminated by a diffuse X-ray corona (in yellow, its actual shape is arbitrary) of various physical extent. The size of the corona is effectively parametrized by $M_\mathrm{i} = \cos{\delta_\textrm{i,limit}}$, which is the upper limit of $\cos{\delta_\textrm{i}}$ up to which we uniformly integrate the impinging radiation represented by the dashed arrows. The blue arrow indicates the viewing inclination $i$ of the observer with respect to the disc normal, i.e. the emission angle $\delta_\textrm{e}$ of the local reflection situated in the disc plane. The local reflection is precomputed and integrated uniformly also across the azimuthal directions $\Phi_\textrm{e}$.}
	\label{xsstokes_disc_sketch}
\end{figure}

\section{Reflection from a torus}\label{compare_xsstokes}

In this section, we will show the results of the new model {\tt xsstokes\_torus}, representing reflection from a circular torus, and compare it mainly to the MC results extensively discussed in Paper I.

\subsection{Spectra and energy profiles of polarization}

\begin{figure*}
	\includegraphics[width=2.\columnwidth]{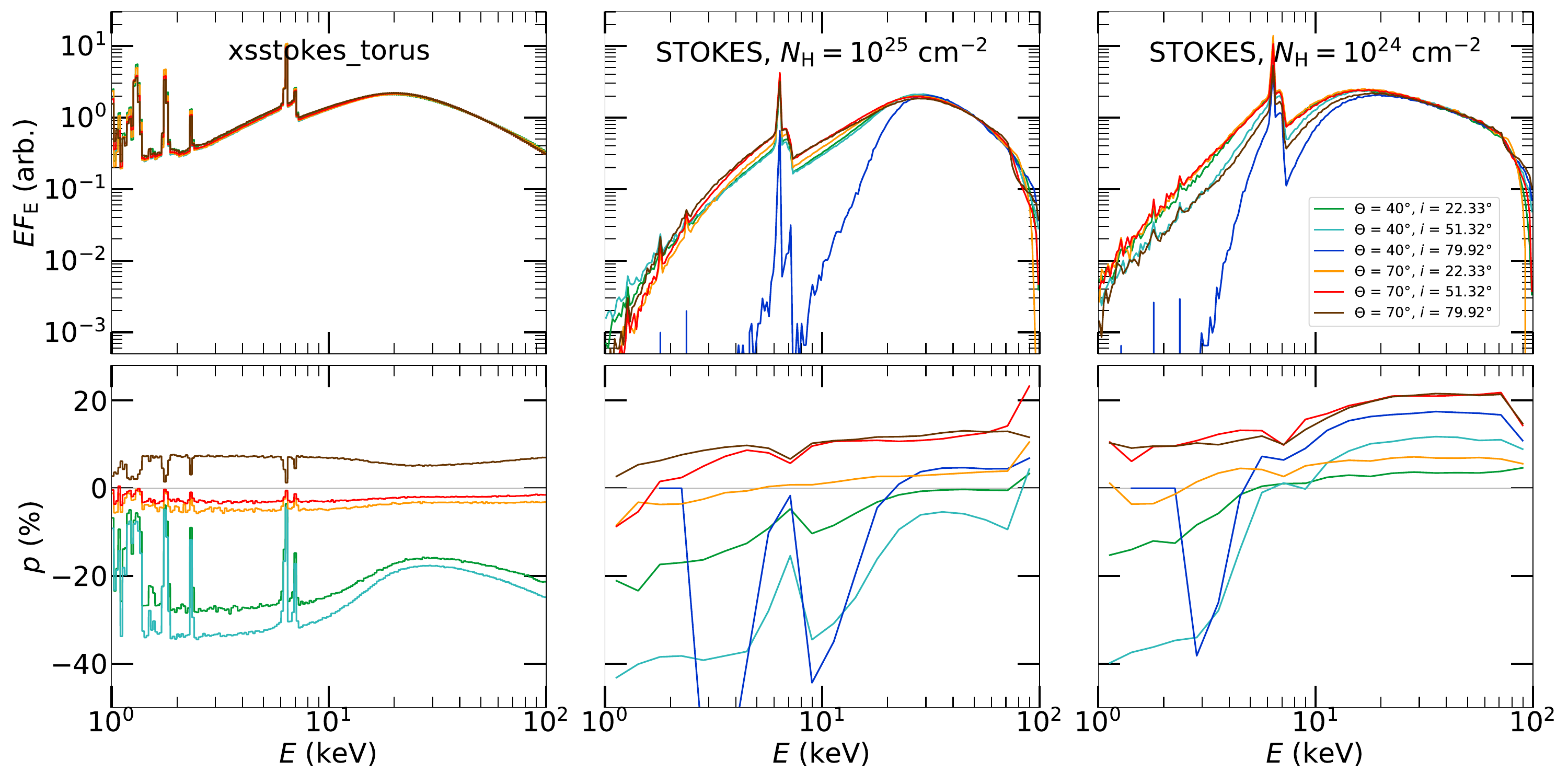}
	\caption{Top: the spectra, $EF_\mathrm{E}$, normalized to value at 50 keV. Bottom: the corresponding polarization degree, $p$, versus energy. We compare the {\tt xsstokes\_torus} results (left) in the same geometry with the results of {\tt STOKES} MC simulations in highly ionized regime for equatorial column densities $N_\mathrm{H} = 10^{25}  \textrm{ cm}^{-2}$ (middle) and $N_\mathrm{H} = 10^{24}  \textrm{ cm}^{-2}$ (right) taken from figure 2 of Paper I. Various inclinations $i$ and half-opening angles $\Theta$ for both type-1 and type-2 observers are shown in the color code. The dark blue curve representing a highly obscured scenario is not shown for the  {\tt xsstokes\_torus} model, because in such case the inner reflecting walls are hidden for the observer in this model. The primary input was set to $\Gamma = 2$ and $p_0 = 2$\% for all displayed results. The detailed parametric dependencies are shown in following figures when integrated in energy, which also helps to reduce the numerical noise in MC simulations.}
	\label{energy_dependent}
\end{figure*}

Figure \ref{energy_dependent} shows the energy-dependence of flux and polarization from both models. The torus in {\tt xsstokes\_torus} has identical meridional profile with the Case C geometry taken from Paper I, to which we compare. Thus, curvature differences and relative size effects are eliminated from the comparisons. Despite much lower numerical noise in {\tt xsstokes\_torus} (only given by the precomputed reprocessing inside the walls, the convergence of the numerical integration is ensured), we see the expected depolarization in spectral lines in both models, regardless the net amplitude and sign of continuum polarization, which is geometry dependent and model dependent.

For the {\tt STOKES} MC simulations plotted for the same free electron density, the energy profile and overall amplitude and sign of the continuum polarization is driven by the relative presence of absorption (cf. the energy dependence of spectra and polarization in all curves on the center and right panels of Figure \ref{energy_dependent} and also the differences in between these two panels). It was discussed in Paper I that more absorption (higher $N_\mathrm{H}$ and/or lower X-ray energies) in partially transparent models causes higher contribution of perpendicular polarization due to dominant scattering rather in the meridional plane, as unabsorbed photons need to travel on average rather near the rotation axis before being scattered and reaching the observer. More transparency allows photons passing through the sides, where scattering rather in the equatorial plane contributes to parallel polarization, which dilutes the perpendicular polarization or even dominates. Details of such competition in the total view depends on geometrical parameters. The obscuring material is generally more transparent (hence parallelly polarized) at high energies than at low energies.

The partial ionization is treated in the MC simulations simplistically by only uniformly adding free electrons on top of a homogeneous distribution of neutral atoms. This results in correct energy-independent amplification of reflected X-ray flux for higher free electron density due to uniformly larger albedo at all energies via scattering. But regarding the energy-dependent absorption and the spectral shape, the model inevitably over-estimates absorption in the soft X-rays for higher ionization cases presented, where we would expect the total absorption opacity to be driven by all ions in some complex ionization structure depending on position inside the obscuring medium. Regardless of a particular geometry, especially for high ionization such simplified treatment of partial ionization results in lack of unabsorbed soft photons for both reflection and transmission (higher soft polarization for both perpendicularly and parallelly polarized contributions), as compared to even more realistic scenarios. One should have this in mind not just for understanding the diversity of results between the two different models, but also before claiming the compared MC models more realistic in all aspects than {\tt xsstokes\_torus}. In the soft X-rays, the {\tt xsstokes\_torus} model produces better predictions for less ionized and denser obscurers than the MC models presented due to its proper treatment of partial ionization and \textit{near} neutrality of the walls (with all ions considered for reprocessing, not just free electrons on top of neutral atoms). For more ionized and/or less dense obscurers, the MC models are more appropriate with the caveat of over-estimated typically negative total polarization in the soft X-rays, especially for high ionization. The correct limit of full ionization with {\tt STOKES} with absorbers removed from the MC model will be additionally discussed below for various densities of free electrons.

The energy profile of the continuum polarization from {\tt xsstokes\_torus} is easier to explain, as we impose no transparency and only study the reflected radiation from the torus surface. The resulting polarization profile with energy follows that of the local reflection tables from \cite{Podgorny2021}: multiple scattering in the Compton hump near 20 keV dilutes the polarization as photons on average scatter through more directions than elsewhere. The increase of polarization at energies above the Compton hump is given by lower number of scatterings of those photons that reached the observer at such energies from the original power-law distribution (each high-energy photon down-shifts in energy at each Compton scattering event as opposed to elastic Thomson scattering). The increase of polarization at energies below the Compton hump is given by enhanced photoelectric absorption at lower energies that effectively traps photons that would scatter multiple times before reaching the observer, which would have led into more diversity of scattering directions in the net view, hence depolarization. The direction and total amplitude of such reflection-induced polarization profile is then given by the geometrical parameters (see below). From the soft spectra it is visible that even the most neutral reflection tables from \cite{Podgorny2021} that we adopted in {\tt xsstokes\_torus} do not represent completely photo-electrically absorbed soft X-rays that we see in the {\tt STOKES} MC simulations. This results in leveling of the continuum polarization fraction with energy at about 2--5 keV in {\tt xsstokes\_torus}. We see such plateau of polarization with energy for any sign of resulting polarization. The sign does not change with energy for any combination of model parameters, as we have only reflection in this model, not diluted by transmitted photons with different polarization orientation towards hard X-rays. If the absorbing atoms were fully neutral in {\tt xsstokes\_torus}, the photo-electric absorption cross-section behaving as $\sim E^{-3}$ would cause a similarly increasing polarization from 5 keV towards the lower energies (but again in both positive and negative resulting polarization, as the no-transparency condition holds compared to the MC results).

\subsection{Continuum polarization dependence on geometry, partial transparency and ionization}

\begin{figure*}
	\includegraphics[width=2.\columnwidth]{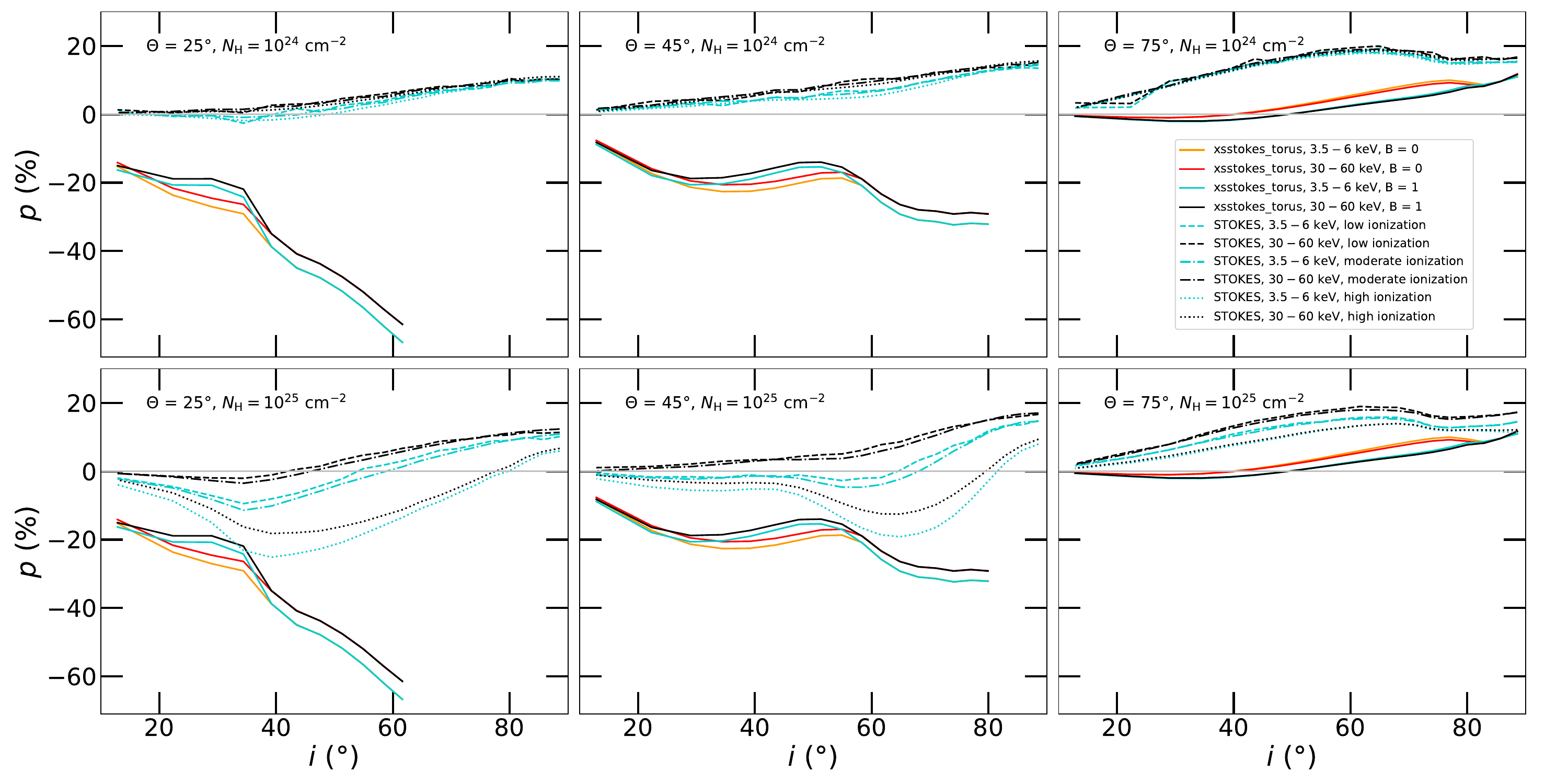}
	\caption{Comparison of the {\tt STOKES} MC simulations from figures A3 and A4 of Paper I (with different line styles assigned to different ionization levels) with the {\tt xsstokes\_torus} results (solid). We plot the polarization degree $p$ versus inclination $i$ for $\Theta = 25\degr$ (left), $\Theta = 45\degr$ (middle), and $\Theta = 75\degr$ (right). We show the results integrated in 3.5--6 keV (blue) and 30--60 keV (black) and for $N_\mathrm{H} = 10^{24}  \textrm{ cm}^{-2}$ in {\tt STOKES} (top) and $N_\mathrm{H} = 10^{25}  \textrm{ cm}^{-2}$ in {\tt STOKES} (bottom). The results of the {\tt xsstokes\_torus} model that do not take into account reflection below the equator (parameter $B = 0$) are additionally shown in 3.5--6 keV (yellow) and 30--60 keV (red). Otherwise the results of the {\tt xsstokes\_torus} model are displayed for the entire illuminated torus surface (parameter $B = 1$) and the {\tt STOKES} results always allow photons to travel anywhere inside the 3D scattering structure, even escaping and reprocessing multiple times. The primary input was set to $\Gamma = 2$ and $p_0 = 2$\% for all displayed cases. The {\tt xsstokes\_torus} simulations do not cover the entire space in $i$ and $\Theta$, because for high $i$ and small $\Theta$, the reflecting area is obscured.}
	\label{CvsD_pmue}
\end{figure*}

\begin{figure*}
	\includegraphics[width=2.\columnwidth]{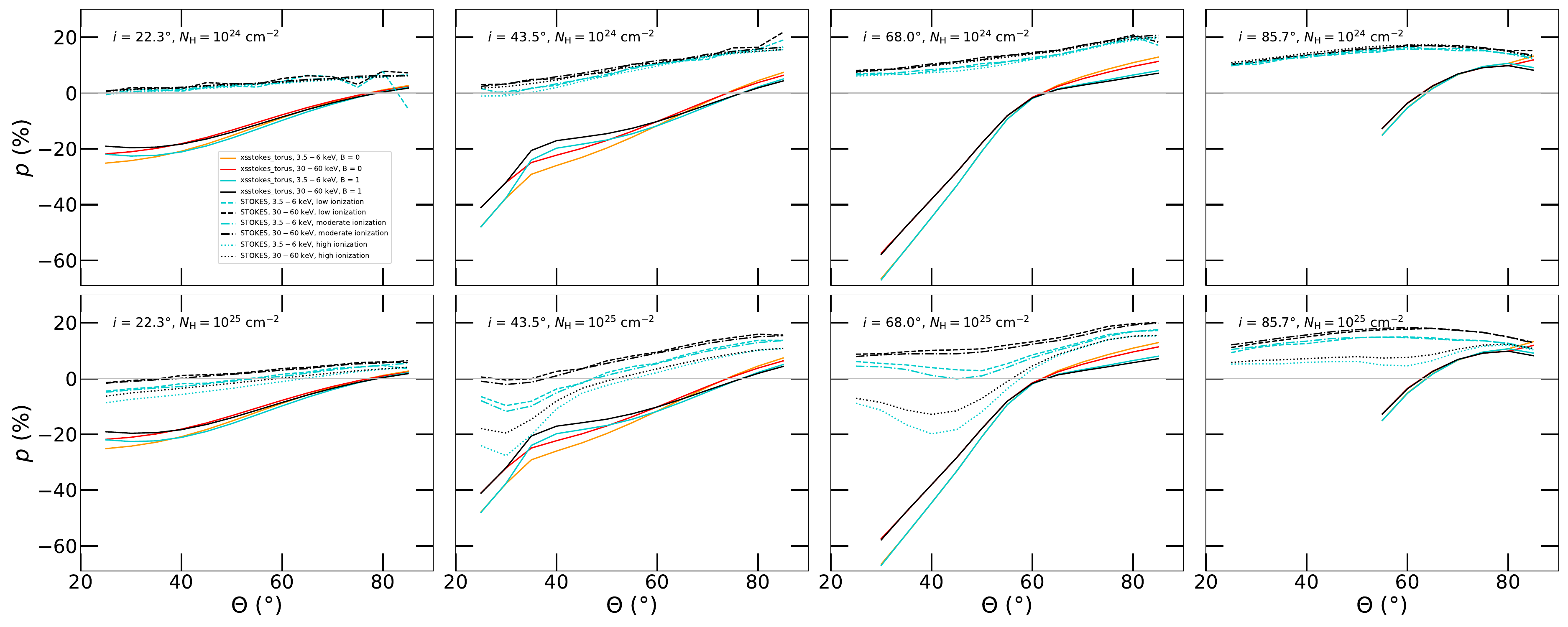}
	\caption{The same as in Figure \ref{CvsD_pmue}, but energy-averaged polarization degree, $p$, versus half-opening angle $\Theta$ is shown for $i = 22.3\degr$, $i = 43.3\degr$, $i = 68.0\degr$, and $i = 85.7\degr$ (left to right).}
	\label{CvsD_ptheta}
\end{figure*}

To disentangle the effects of partial transparency and ionization in more detail and with respect to the varying torus geometry and observer's inclination, we will proceed by studying energy-integrated results in particular bands in the absence of spectral lines. We show in Figures \ref{CvsD_pmue} and \ref{CvsD_ptheta} the same integrated polarization in 3.5--6 keV and 30--60 keV as in figures A3 and A4 in Paper I, but with the {\tt xsstokes\_torus} added. We also plot the {\tt xsstokes\_torus} results without taking the bottom half-space of the reflecting torus into account (mode $B = 0$ in the model, expecting some optically thick material present inside the torus cavity, see Appendix \ref{torus_calculus} for details). This is only to show that the contribution from the illuminated areas below the equator is important for low inclinations when they are observable and the overall change in predicted polarization is not very significant. The new model is in both variants compared to the MC simulations for low, moderate and high partial ionization levels (see Paper I for details for the values of free electron densities relative to neutral atoms). The highest resemblance between the two models generally happens for the high ionization and high densities of the MC simulations, which may seem surprising at first, as the results should match for low ionization rather. The key model difference is the remaining partial transparency in {\tt STOKES} that dilutes the reflection-induced perpendicular polarization even for $N_\mathrm{H} = 10^{25}  \textrm{ cm}^{-2}$. If we had used even higher equatorial column density in the MC simulations, all curves (for any ionization level presented) would move towards more perpendicularly polarized results (i.e. towards the {\tt xsstokes\_torus} model representing the opaque limit), as the grazing angle column densities for the surface layers would gain sufficiently high optical depths, which is critical for the effective reduction of symmetry in scattering directions and the perpendicular polarization gain. Such discrepancy in polarization due to partial transparency is naturally larger for higher $i$ and lower $\Theta$. With 3D travelling allowed inside the torus and self-irradiation included in the MC models, we notice high parallel polarization contribution for high inclinations and low half-opening angles even for $N_\mathrm{H} = 10^{25}  \textrm{ cm}^{-2}$ and the highly ionized case. This is because in such highly obscured configurations, the photons are still allowed to scatter multiple times in the upper regions of the torus before escaping to the observer, which defines the dominant plane of last scattering events and the prevailing parallel polarization orientation. The upper panels of Figures \ref{CvsD_pmue} and \ref{CvsD_ptheta} show almost no difference in ionization level, because the densities of the absorbers are for all ionization cases too low to enforce a more restricted global geometry (causing obscuration and beaming) and the torus is optically thin. The direction of change of polarization towards more negatively polarized results with higher ionization in MC simulations in the bottom panels of Figures \ref{CvsD_pmue} and \ref{CvsD_ptheta} follows from the competition between reflected and transmitted X-rays in the MC model. For higher ionization, there are more reflected photons from the surface due to high relative fraction of free electrons with respect to the absorbers and the back-side reflection dominates over transmission, which induces rather perpendicularly polarized photons (for low and intermediate $\Theta$; at higher $\Theta$ reflection from the left and right sides of the torus from observer's perspective occupies greater projected area contribution, which again induces parallel polarization).

This effect of relative flux contribution from different projected areas of the scattering medium in the partially transparent model is much more important than the change in polarization due to the average number of scatterings per photon. If the partial transparency was forbidden in the pure MC model, then the change of free electron density relative to the density of absorbers would only affect the average number of scatterings allowed when the photons enter the surface and escape. Such impact of changing ionization on polarization via average number of scatterings is smaller and in the reverse direction for net negative polarization (more scattering orders for higher ionization mean lower polarization for any resulting sign). And it is in fact seen in the comparison of the two different energy bands for the {\tt xsstokes\_torus} model, which has the no-transparency condition and differentiates only the number of scatterings per energy band due to change of absorption to scattering opacity ratio with energy. It is also supported by the additional single-scattering results discussed below.

\subsection{Imaging results}

\begin{figure*}
    \centering
    \includegraphics[width=2.\columnwidth]{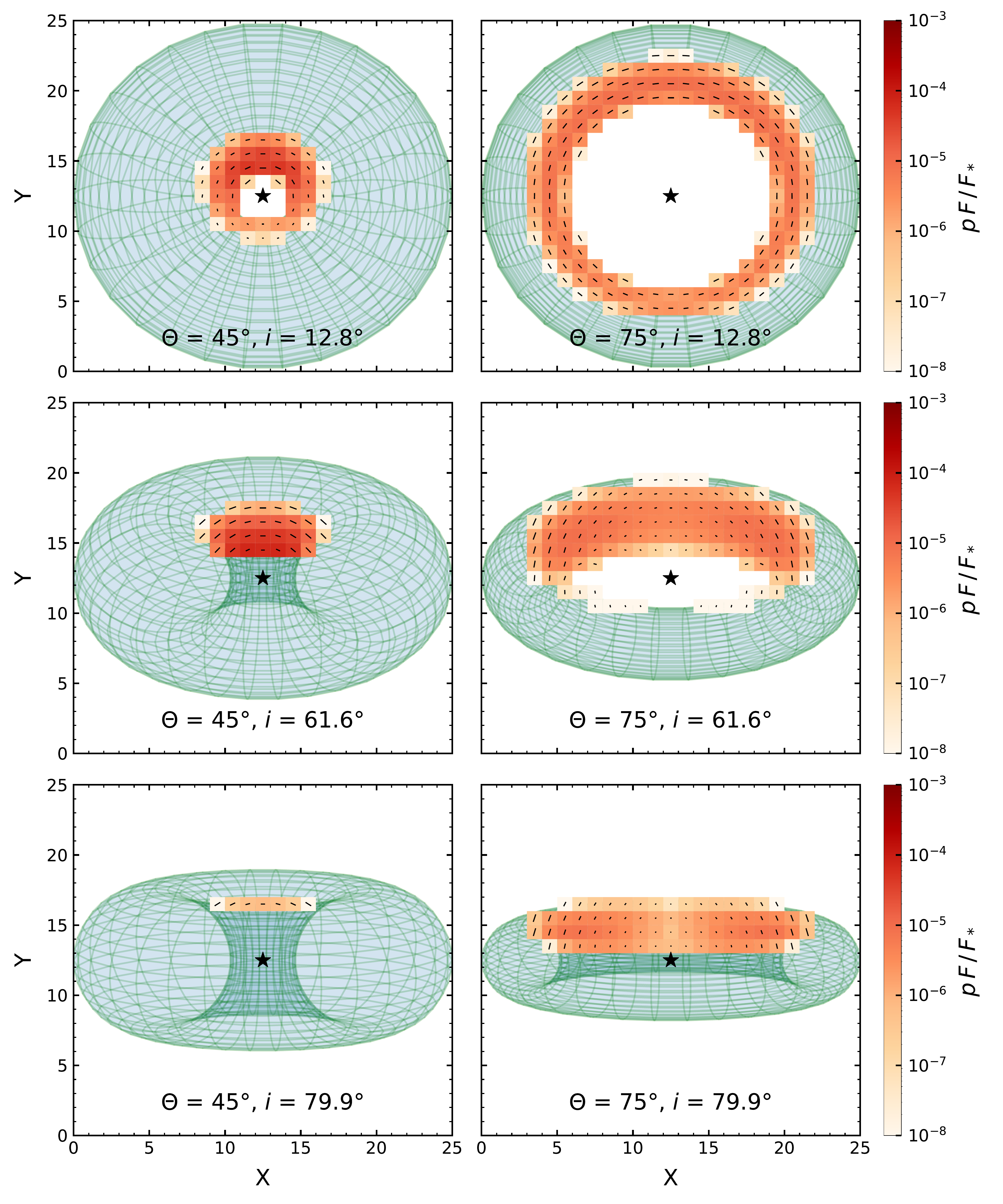}
    \caption{The images in cartesian coordinates $X$ and $Y$ on the sky of the projected torus (with its rotation axis aligned with $Y$) from the {\tt torus\_integrator}, which precomputes the tabular dependencies of the {\tt xsstokes\_torus} model. The location of the central isotropic source is marked by the black star. In the color coded pixels we show the projected contributions of the torus surface that are both illuminated by the central source and directly observable, with the color corresponding to the $pF/F_\mathrm{*}$ quantity, i.e. the polarized flux divided by the flux of the source towards the inclined observer, if the source was unobscured. Each of the contributing projected surface areas is additionally marked with a black bar with length proportional to the locally induced linear polarization degree and tilt corresponding to the locally induced linear polarization direction. Vertically aligned bars represent the polarization orientation aligned with the system rotation axis. The six panels represent different combinations of the torus half-opening angles $\Theta$ and observer's inclinations $i$. The results are integrated in 3.5--6 keV.}
    \label{imaging_xsstokes}
\end{figure*}

\begin{figure*}
    \centering
    \includegraphics[width=2.\columnwidth]{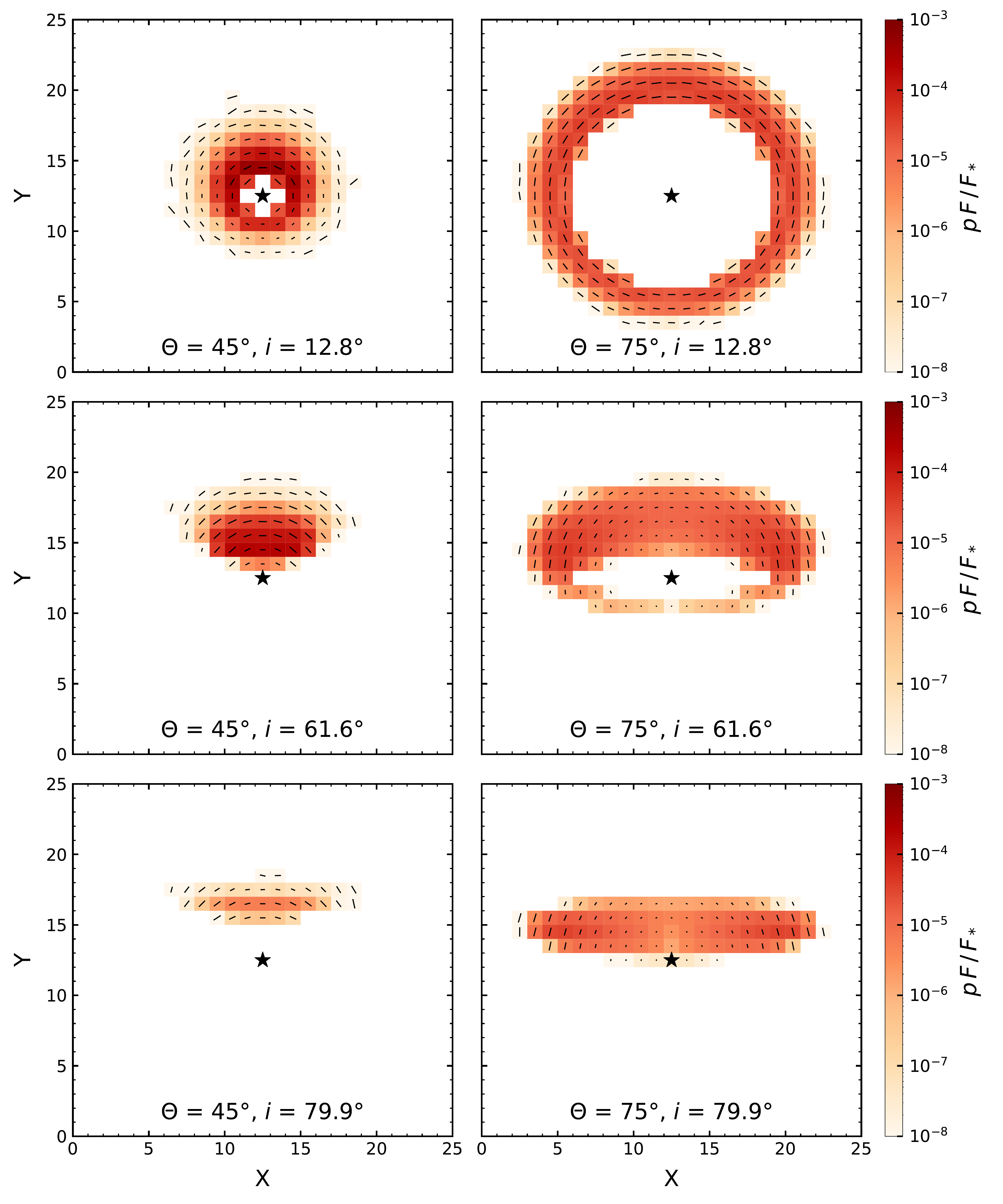}
    \caption{The same as in Figure \ref{imaging_xsstokes}, but the data are taken from the {\tt STOKES} MC simulation for the highly ionized case with $N_\textrm{H} = 10^{25}  \textrm{ cm}^{-2}$ and $\tau_\textrm{e} = 7$. Pixels with low number of photons show high numerical noise and we do not provide the polarization bar for clarity. In this regime, we obtain the highest resemblance with the {\tt xsstokes\_torus} model, because the torus walls show low transparency and the reflection is enhanced. Depending on the viewing angle and torus half-opening angle, parallell or orthogonal polarization direction dominates in the spatially integrated view.}
    \label{imaging_STOKES_NH25}
\end{figure*}

\begin{figure*}
    \centering
    \includegraphics[width=2.\columnwidth]{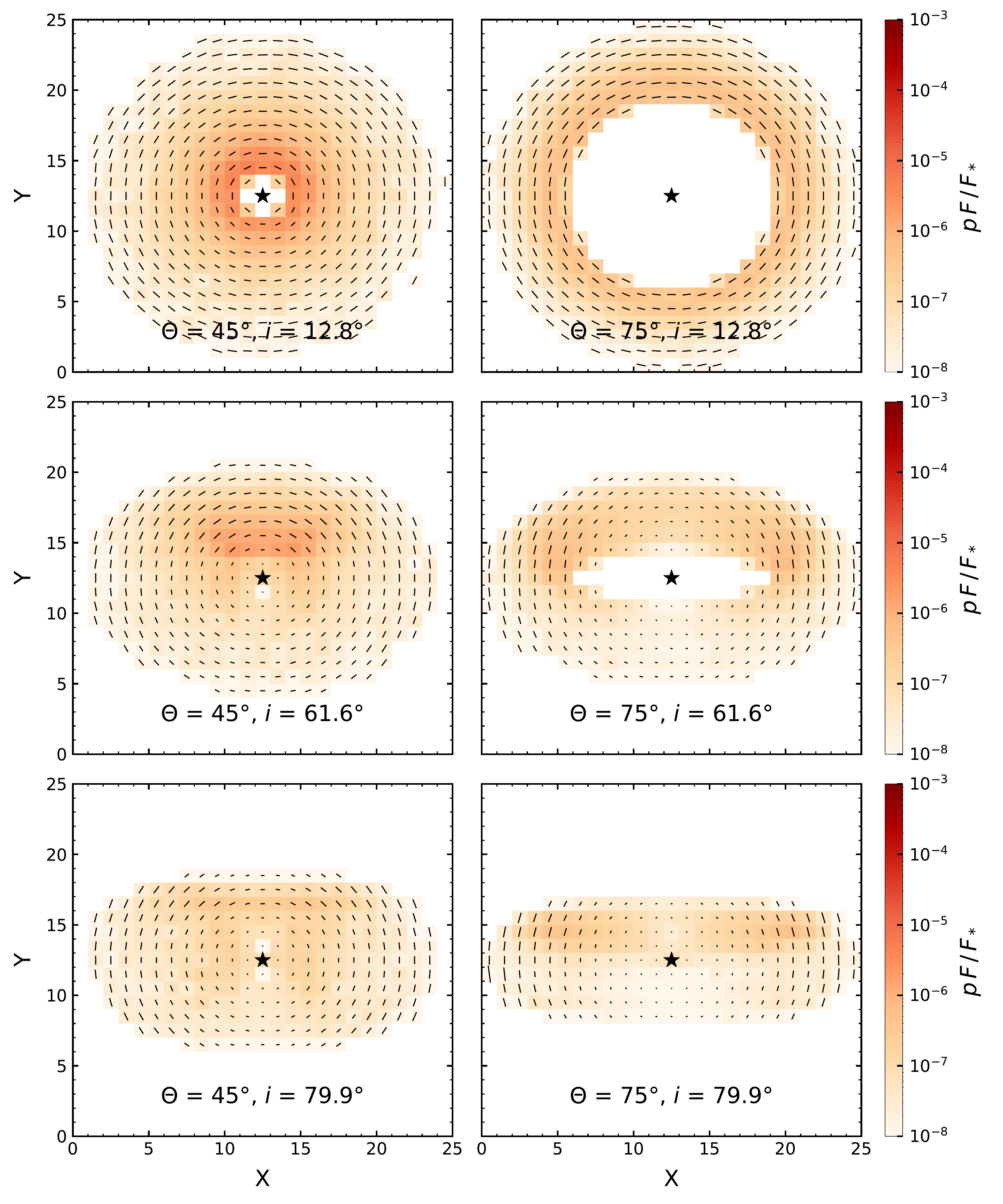}
    \caption{The same as in Figure \ref{imaging_xsstokes}, but the data are taken from the {\tt STOKES} MC simulation for the low ionized case with $N_\textrm{H} = 10^{24}  \textrm{ cm}^{-2}$ and $\tau_\textrm{e} = 0.007$. In this regime, the entire scattering region is highly transparent and we observe predominantly parallel polarization with the system axis, because most photons arriving to the observer scatter in the equatorial plane.}
    \label{imaging_STOKES_NH24}
\end{figure*}

In order to strengthen the above analysis, we show in Figure \ref{imaging_xsstokes} the projected torus images for various inclinations and half-opening angles from the computations that underlie the {\tt xsstokes\_torus} model. In Figures \ref{imaging_STOKES_NH25} and \ref{imaging_STOKES_NH24} are the corresponding MC examples for high neutral density with high ionization and low neutral density with low ionization, respectively, that show the biggest integrated polarization state differences in the selected MC examples. With high density of both neutral species and free electrons in the MC model we obtain high perpendicular polarization from the opposite side of the reflecting inner torus surface, which provides relatively strong flux contribution with respect to angular distributions of photons per visible surface area. For high half-opening angles and all viewing angles, the reflection from the left and right sides from observer's perspective occupies larger projected areas relative to the back side producing perpendicularly polarized signal. Thus, for high half-opening angles even for the {\tt xsstokes\_torus} model the integrated results are parallelly polarized, although we observe lower total polarization fraction by more than half compared to the integrated MC results due to the difference in partial transparency. The higher the half-opening angle, the more equatorially distributed is the scattering medium with respect to the central source. In the limit of high half-opening angles, the torus is representing a disc ring and should match its polarization properties (see Section \ref{distant_disc}).

\subsection{Special cases of full ionization and full neutrality}

\begin{figure*}
    \centering
    \includegraphics[width=2.\columnwidth]{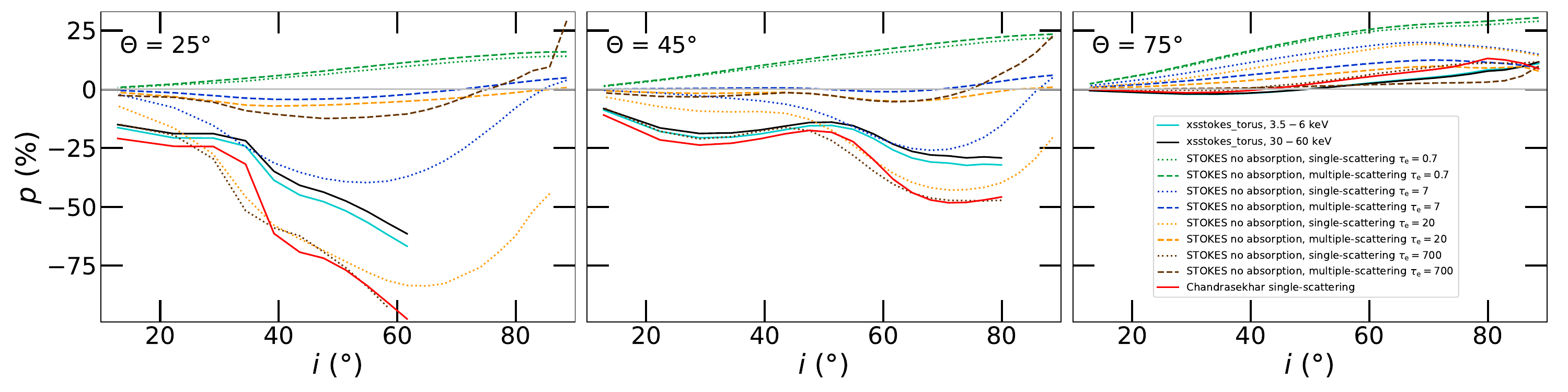}
    \caption{The polarization degree $p$ versus inclination $i$ for $\Theta = 25\degr$ (left), $\Theta = 45\degr$ (middle), and $\Theta = 75\degr$ (right), as in Figure \ref{CvsD_pmue}. We compare the {\tt STOKES} MC simulations computed with homogeneously distributed free electrons only (different line colors are assigned to different equatorial Thomson optical depths $\tau_\mathrm{e}$) and the {\tt xsstokes\_torus} results integrated in 3.5--6 keV (solid light blue) and 30--60 keV (solid black). The {\tt STOKES} simulations were computed with multiple scatterings allowed (dashed lines, the full ionization model) and with single scatterings (dotted lines), both integrated in 3.5--6 keV, but note that the polarization produced without absorbers in this range is energy independent. The primary input was set to $\Gamma = 2$ and $p_0 = 2$\% for all displayed cases. We also analogically provide the results of the {\tt xsstokes\_torus} model when a more simple energy-independent prescription is used for local reflection: the Chandrasekhar's single-scattering analytical formulae (red).}
    \label{additional_simulations}
\end{figure*}

Next, we removed the absorbers from the {\tt STOKES} MC simulations and made additional simulations with free electrons only for a) multiple scatterings allowed (the correct full ionization limit) and b) for single-scattering results (the fully neutral limit reached for high electron densities). Because in the toroidal geometries the line-of-sight density near grazing inclinations is significantly lower than the equatorial density (unlike e.g. wedge-shaped geometries for obscurers), high equatorial electron optical depths are needed in the MC simulations to ensure optically thick surface layers seen under grazing inclination angles. The results for both multiple scatterings and single-scattered photons are in Figure \ref{additional_simulations} for various equatorial Thomson optical depths $\tau_\mathrm{e} = 2r\sigma_\mathrm{T}n_\mathrm{e}$ (where $\sigma_\mathrm{T}$ is Thomson cross-section and $n_\mathrm{e}$ is the homogenous electron number density), compared directly to the {\tt xsstokes\_torus} model. Indeed, the higher the equatorial $\tau_\mathrm{e}$, the more perpendicular polarization contribution we obtain from the MC simulations, as the scattering geometry is reduced to an optically thick funnel. The single-scattered photons show higher polarization degree than the multiple-scattering results (for any sign of the resulting polarization state that is given by $\Theta$ and $i$, as this is a general effect related to the level of asymmetry in scattering directions in the medium). Only for the lowest $\tau_\mathrm{e}$ there is almost no difference\footnote{Even when multiple scatterings are allowed, the photons scatter on average once before reaching the observer. Photons that did not scatter even once (direct primary radiation) are not registered in the simulation.} and the simulations resemble the optically thin pure-scattering polarization dependencies on inclination in the literature for similar geometries \citep{Sunyaev1985, Tomaru2024}.

\begin{figure*}
    \centering
    \includegraphics[width=2.\columnwidth]{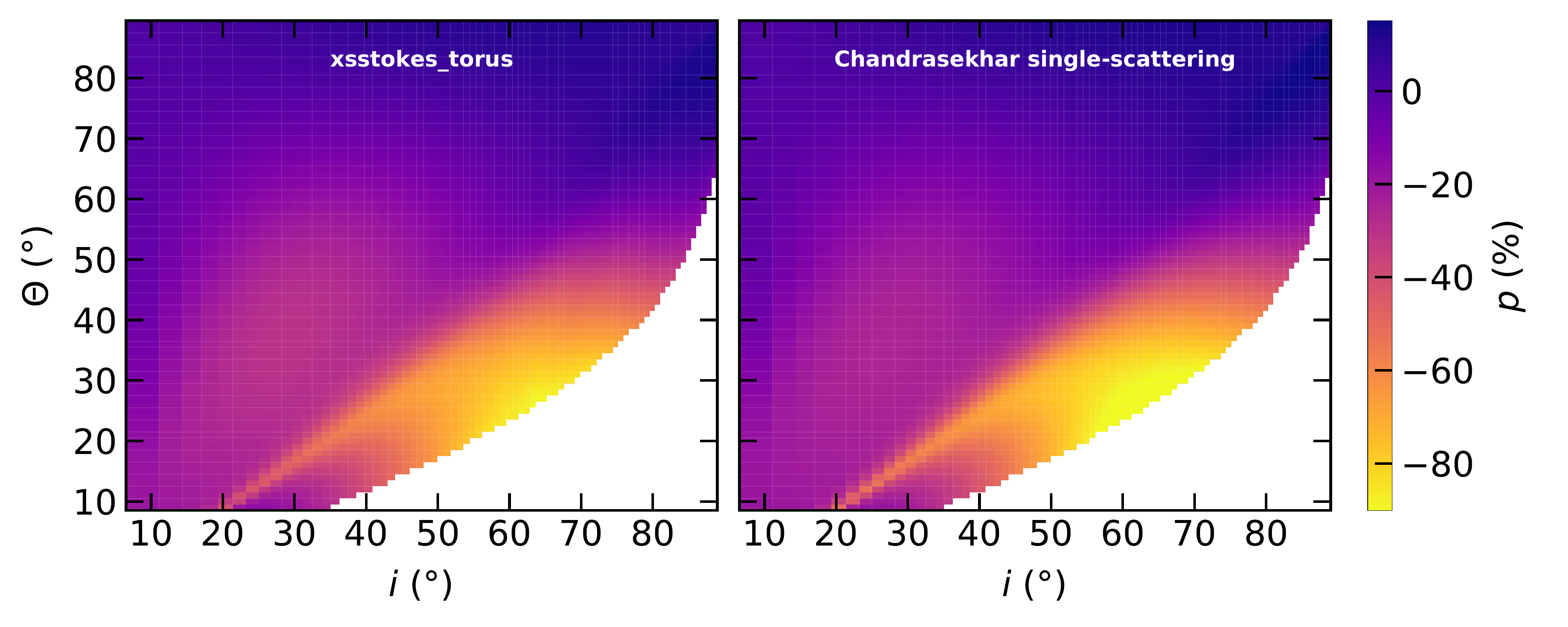}
    \caption{The polarization degree $p$ versus observer's inclination $i$ versus torus half-opening angle $\Theta$. We provide the full parameter space from the {\tt xsstokes\_torus} model, using the local reflection tables from \citet{Podgorny2021} in 3.5--6 keV (left) and when a simple energy-independent prescription from the Chandrasekhar's single-scattering formulae is taken instead (right). The polarization pattern with changing geometry matches, only the obtained polarization fraction is different, which is more observable in Figure \ref{additional_simulations} that provides a few slices in $\Theta$ of these heatmaps. The results are provided for $\Gamma = 2$ and 2\% parallelly polarized primary. The values of $\Theta(i)$ below the $\Theta_\mathrm{limit}(i)$ are directly not observable, thus marked in white. See Appendices \ref{technical_details} and \ref{torus_calculus} for more information.}
    \label{fullspace_vs_Chandra}
\end{figure*}

The high-density single-scattering results from the MC simulations allow the scattering to occur in very small volumes, practically on the surface of the torus, thus best matching the {\tt xsstokes\_torus} model. However, the {\tt xsstokes\_torus} takes into account some energy-dependent level of average scattering orders thanks to its precomputed ionization structure. To cross-validate the models in terms of geometry, we replaced the local reflection in {\tt xsstokes\_torus} with energy-independent Chandrasekhar's analytical formulae for reflection gained via Rayleigh single-scattering \citep{Chandrasekhar1960}. Finally, these curves (in red in Figure \ref{additional_simulations}) reproduce the high-density single-scattering MC simulations reasonably well.\footnote{The remaining differences between the solid red and dotted brown curves in Figure \ref{additional_simulations} could be due to the fact that {\tt STOKES} in the adopted setup adds photons symmetrically with respect to the equatorial plane to reduce MC numerical noise and computational times. Further investigations of this effect remains to be done.} Their shape also well follows the original {\tt xsstokes\_torus} model with even higher perpendicular polarization induced due to the neglecting of higher scattering orders. To prove that the general behavior of polarization is alike for all $i$ and $\Theta$ we plot in Figure \ref{fullspace_vs_Chandra} the entire parameter space in \{$i$,\,$\Theta$\} with {\tt xsstokes\_torus} when using both local reflection prescriptions. Plotting the full geometrical parameter space is also useful for comparisons with other models in the literature that assume different than toroidal obscurers for the (nearly) neutral high-density conditions. We obtain similar polarization dependencies on geometry as models in \cite{Ursini2023} and \cite{Veledina2023} that both adopted a reflecting cone.
  
\subsection{Effects of changing primary radiation properties}

\begin{figure*}
	\includegraphics[width=2.\columnwidth]{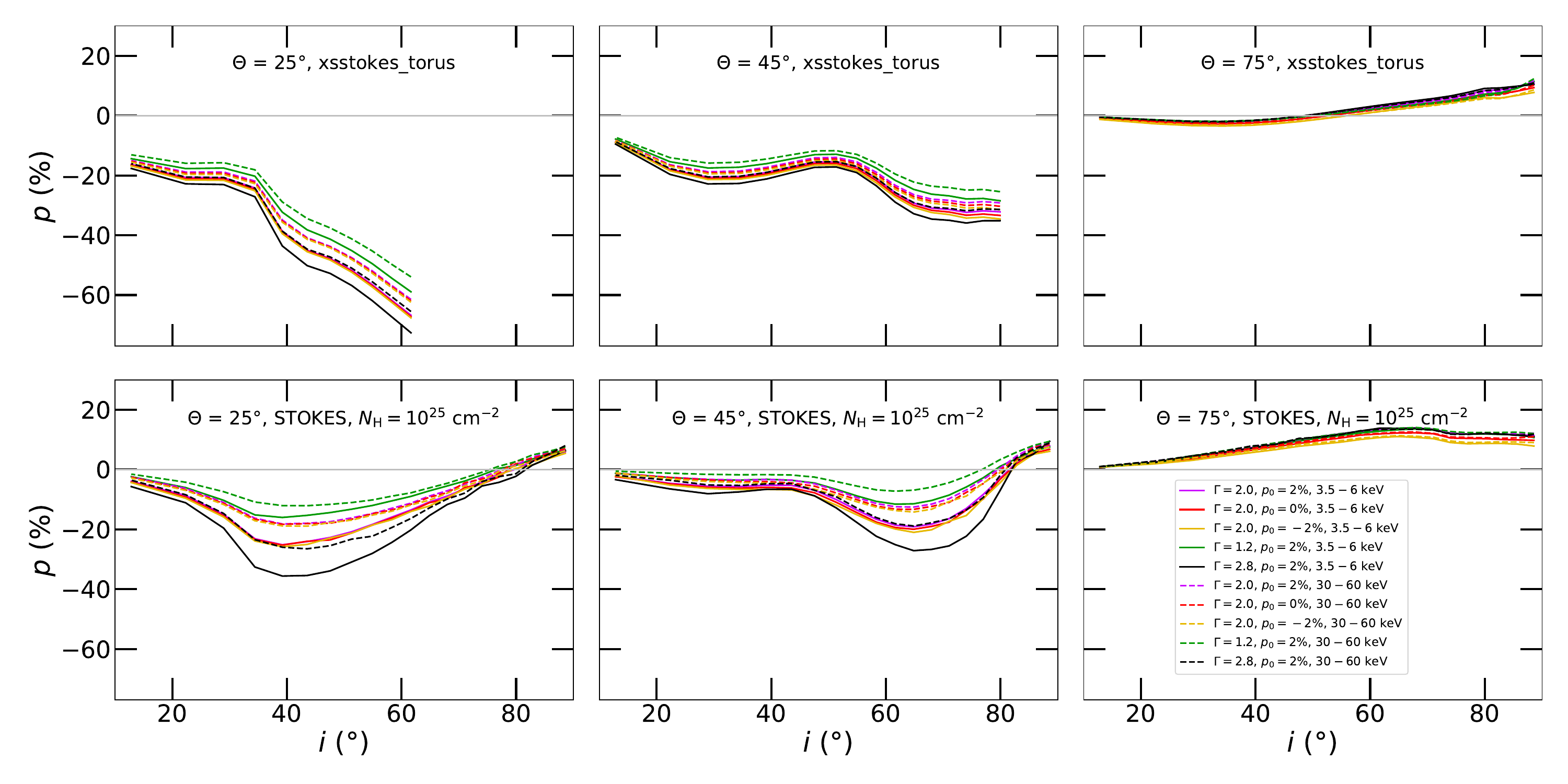}
	\caption{The {\tt xsstokes\_torus} results (top) for different cases of primary radiation, compared in the same way with the {\tt STOKES} MC simulations (bottom) for $N_\mathrm{H} = 10^{25}  \textrm{ cm}^{-2}$ and high ionization taken from figure 10 in Paper I. We plot the energy-averaged polarization degree, $p$, versus inclination $i$ for $\Theta = 25\degr$ (left), $\Theta = 45\degr$ (middle), and $\Theta = 75\degr$ (right). We show the results in 3.5--6 keV (solid) and 30--60 keV (dashed). The magenta curves correspond to the 2\% parallelly polarized primary with $\Gamma = 2.0$. The red and yellow correspond to the unpolarized and 2\% perpendicularly polarized primary with $\Gamma = 2.0$, respectively. The green and black curves correspond to the 2\% parallelly polarized primary with $\Gamma = 1.2$ and $\Gamma = 2.8$, respectively.}
	\label{D_primary}
\end{figure*}

Lastly, let us compare the effects of changing primary radiation properties. Figure \ref{D_primary} shows the same dependencies on changing primary polarization and $\Gamma$ for {\tt xsstokes\_torus} and for the full MC simulation with absorbers (to provide the MC simulations we select a few cases from figure 10 of Paper I). The effects on polarization have comparable extent and the same qualitative behavior. This also cross-validates the new model. The dependency of the net polarization on changing primary polarization state is natural from the scattering theory (see Appendix \ref{incident_polarization}) and we observe similar results for reflection from accretion discs (see Section \ref{distant_disc} or the results in \cite{Podgorny2023}). The dependency of the net polarization on changing $\Gamma$ is given by different weighting of the spectrum on different energies.

\section{Reflection from a faraway disc}\label{distant_disc}

In this section we briefly discuss the results of the {\tt xsstokes\_disc} routine, suitable for studying X-ray polarization signatures from distant rings of AGN accretion discs without any special- or general-relativistic distortion.

\subsection{Energy dependence of polarization}

\begin{figure*}
	\includegraphics[width=2\columnwidth]{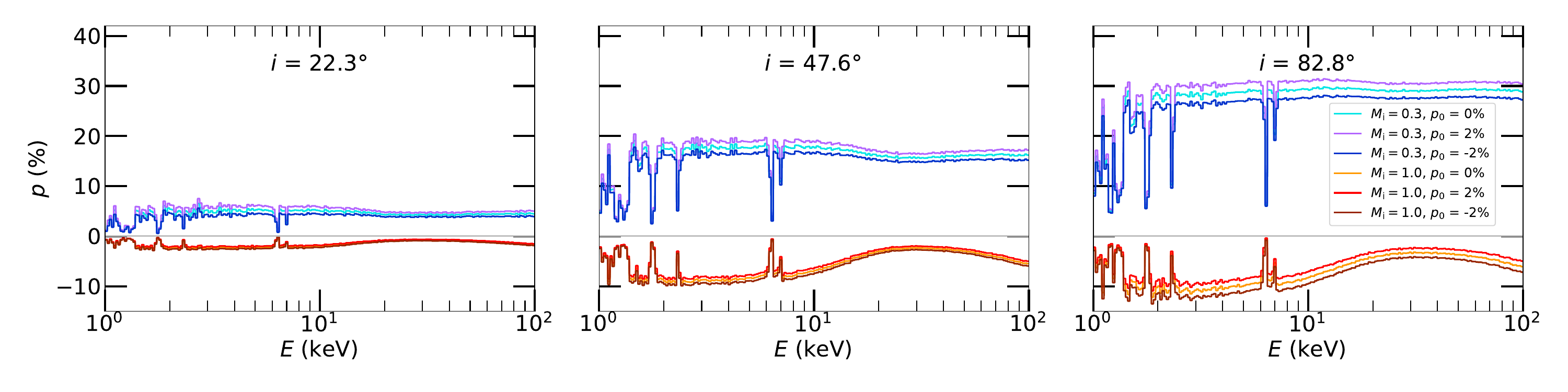}
	\caption{The polarization degree, $p$, versus energy for distant disc reflection computed by {\tt xsstokes\_disc} for $M_\mathrm{i} = 0.3$ (top, parallelly polarized) and $M_\mathrm{i} = 1$ (bottom, perpendicularly polarized). We show the results for various cosines of the observer's inclination: $\mu_\textrm{e} = 0.925$ (left), $\mu_\textrm{e} = 0.675$ (middle), $\mu_\textrm{e} = 0.125$ (right). The color code represents different incident polarizations of the primary power-law with $\Gamma = 2$ for the two cases of $M_\mathrm{i}$: unpolarized, 2\% parallelly polarized, and 2\% perpendicularly polarized.}
	\label{disc_energy_dependent}
\end{figure*}

Figure \ref{disc_energy_dependent} shows the resulting polarization with energy for various model parameters and two values of $M_\mathrm{i}$. The $M_\mathrm{i} = 0.3$ case represents a centrally illuminating corona with fairly low relative height with respect to the disc radial extension. The $M_\mathrm{i} = 1$ case represents a completely diffuse corona around the disc with many photons arriving also vertically along the disc normal. The {\tt xsstokes\_disc} routine for low $M_\mathrm{i}$ ($\lesssim 0.8$) predicts parallel (i.e. positive) polarization, because the reflecting matter is distributed equatorially with respect to the source, which forms the dominant plane of scattering (with polarization direction from the Compton effect perpendicular to it). For e.g. $M_\mathrm{i} = 0.3$, the level of received net polarization fraction can be from 0\% up to $\sim 30$\%, mainly depending on the inclination of the observer, which is the main driver of polarization in this simplistic model.

The spectral lines are depolarized, as expected, and the energy profile of polarization degree follows that of the local reflection tables from \cite{Podgorny2021}, seen already in Figure \ref{energy_dependent} with the {\tt xsstokes\_torus} model. The obtained difference in polarization degree is expected on the output for changing the incident emission by additional small polarized fraction $p_0$. The received emission has \textit{larger} polarization compared to a \textit{parallelly} polarized result computed for unpolarized primary, if the incident polarization is oriented also \textit{parallelly} (effectively adding some fraction of polarization), and is \textit{depolarized} compared to a \textit{parallelly} polarized result computed for unpolarized primary, if the incident polarization is oriented \textit{perpendicularly} (effectively subtracting some fraction of polarization). The opposite is true for perpendicularly polarized results for unpolarized primary. This is consistent with the impact of primary polarization on reflected emission from the accretion disc that was presented for inner regions of the disc including general-relativistic effects in the lamp-post disc-corona system in \cite{Podgorny2023} and with the reflection on the torus (see Figure \ref{D_primary}). We refer to Appendix \ref{incident_polarization} for more theoretical background, showing that for small $p_\mathrm{0}$ the difference in obtained polarization with respect to the results for unpolarized primary scales linearly with $p_\mathrm{0}$ and is not greater than $p_\mathrm{0}$.

\subsection{Dependence of continuum polarization on inclination and coronal size}

\begin{figure}
	\includegraphics[width=1.\columnwidth]{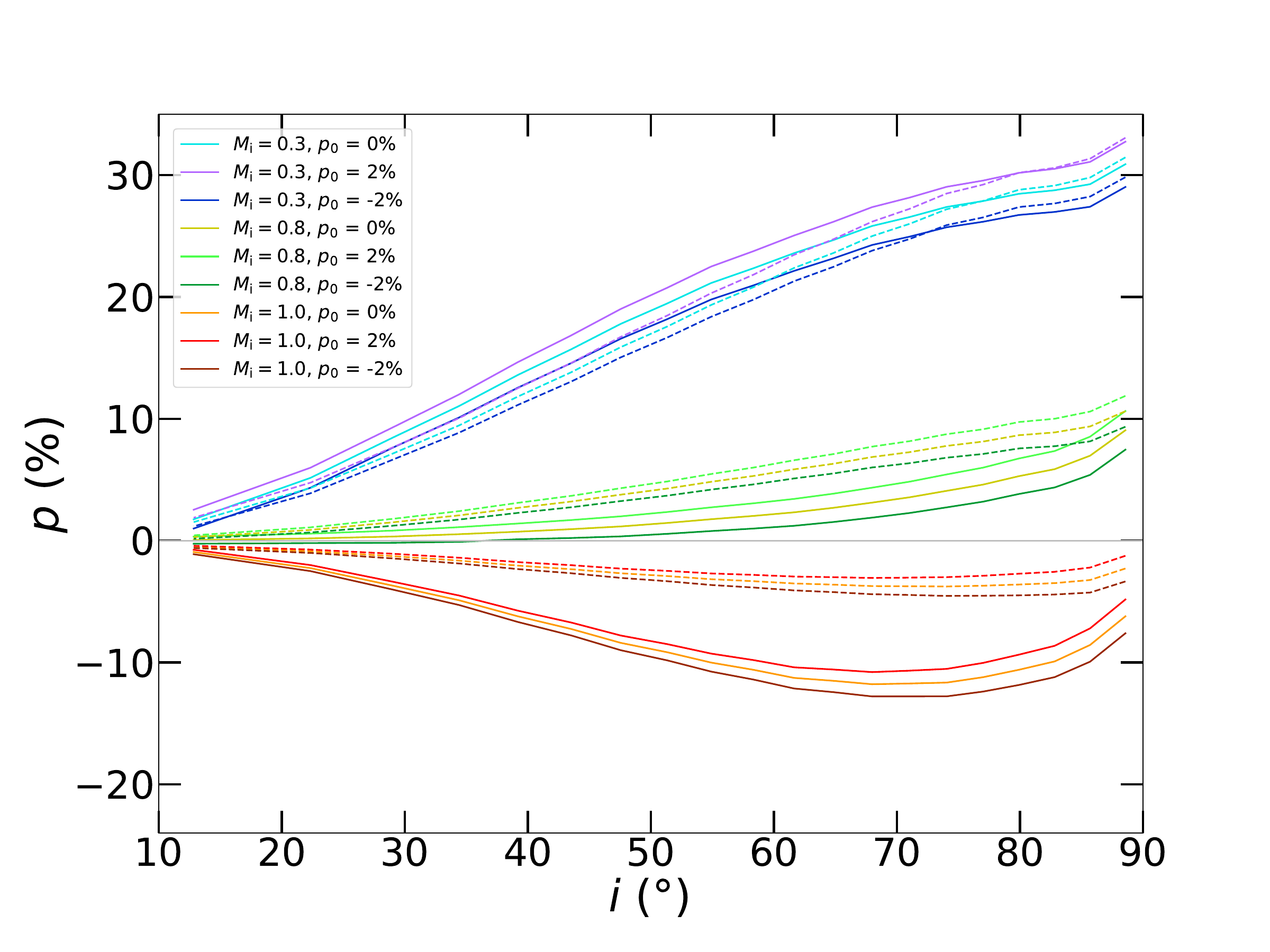}
	\caption{The energy-averaged polarization fraction, $p$, computed with {\tt xsstokes\_disc} in 3.5--6 keV (solid lines) and 30--60 keV (dashed lines) versus observer's inclination $i$ for $M_\mathrm{i} = 0.3$ (top), $M_\mathrm{i} = 0.8$ (middle) and $M_\mathrm{i} = 1$ (bottom). The color code represents different incident polarizations of the primary power-law with $\Gamma = 2$ for the three cases of $M_\mathrm{i}$: unpolarized, 2\% parallelly polarized, and 2\% perpendicularly polarized.}
	\label{disc_mue_dependent}
\end{figure}

To see the inclination dependence of polarization clearly and to examine the dependency on $M_\mathrm{i}$, Figure \ref{disc_mue_dependent} shows the {\tt xsstokes\_disc} results for $M_\mathrm{i} = 0.3$, 0.8 and 1.0, respectively, integrated in two different energy bands with absence of spectral lines. The results for low $M_\mathrm{i}$ are consistent with the high half-opening angle limit seen in the MC simulations and in the {\tt xsstokes\_torus} model, both assuming central isotropic illumination, i.e. a nearly linearly increasing polarization fraction with inclination and parallel polarization direction in all subcases (compare with Figures \ref{CvsD_pmue} and \ref{additional_simulations}, right panels). The reflection from the distant disc provides only a higher polarization fraction due to effectively higher asymmetry of the system compared to the circular tori. The results for low $M_\mathrm{i}$ are consistent with figure 11 from \cite{Ratheesh2021}, which attempted for a similar reflection scenario with the models of \cite{Matt1989, Matt1991}. The geometrical difference with respect to our case was higher isotropy in illumination of the disc, hence receiving about 50\% lower net polarization compared to our results for $M_\mathrm{i} = 0.3$. If we increase the level of isotropy in illumination of the disc by increasing $M_\mathrm{i}$ in our model, the polarization fraction drops monotonically and eventually reaches perpendicular orientation with increasing polarization fraction again towards the extreme values of $M_\mathrm{i} = 1$, as the photons arriving vertically to the disc begin to dominate the resulting polarization. When the corona is fully covering the disc and has non-negligible optical depth, however, we would expect the Comptonized photons that reflect from the disc to re-scatter again in the corona before reaching the observer and change their polarization state once more. Hence, the results for large $M_\mathrm{i}$ are less realistic and show only the theoretical direction of the expected change in polarization by disc reflection when increasing the coronal size above the disc.

\subsection{Comparison with analytical computations}

\begin{figure}
	\includegraphics[width=1.\columnwidth]{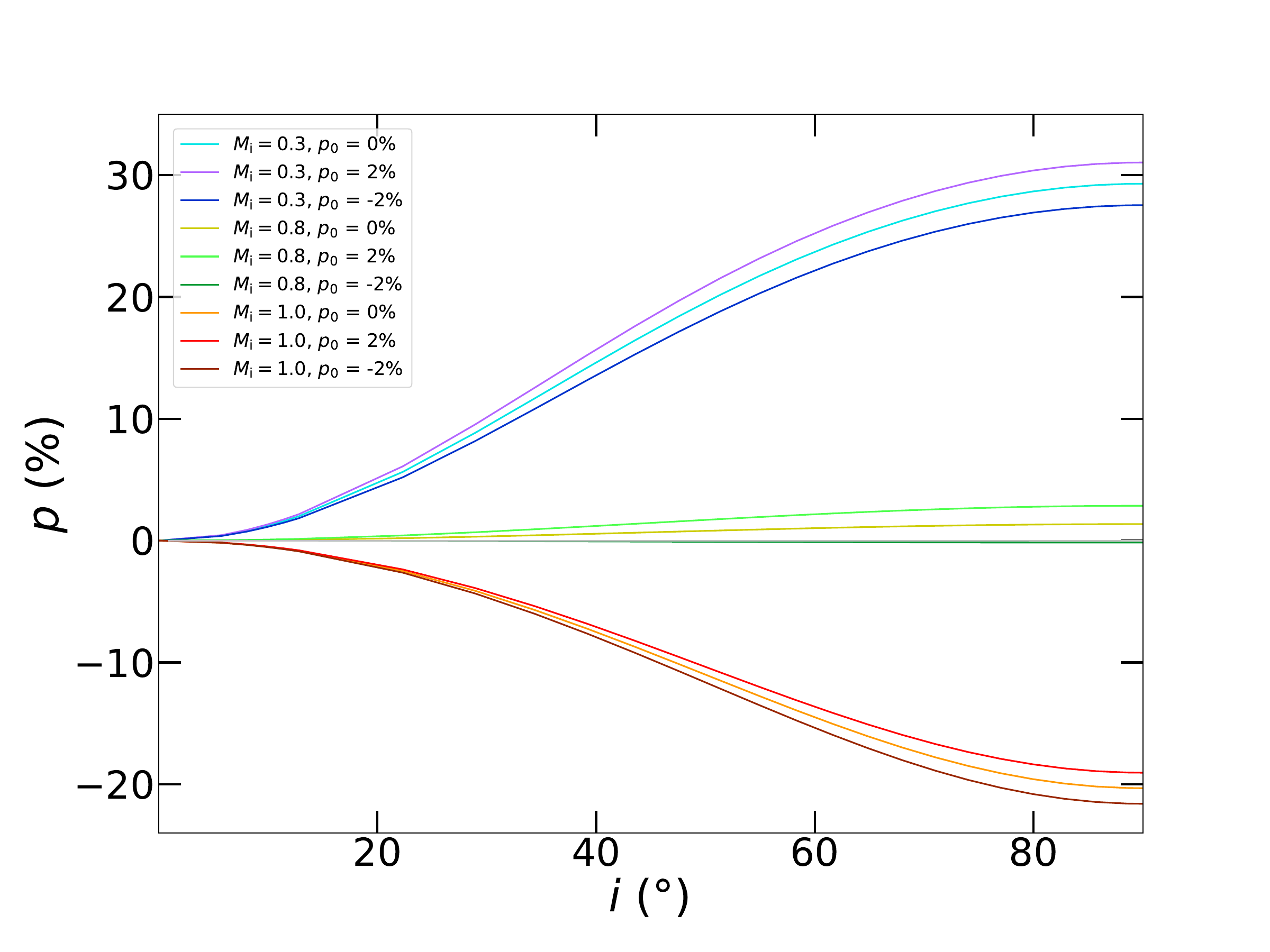}
	\caption{The polarization degree, $p$,  versus observer's inclination $i$ for the energy-independent Chandrasekhar's single-scattering approximation integrated uniformly in all azimuthal emission angles $\Phi_\textrm{e}$ and in selected ranges of cosines of incident angles $0 \leq \mu_\textrm{i} \leq M_\mathrm{i}$, as in the {\tt xsstokes\_disc} model. We display the results for different primary polarizations and $M_\mathrm{i}$ values in the same way as in Figure \ref{disc_mue_dependent} for easy comparison.}
	\label{chandra_comparison}
\end{figure}

In Figure \ref{chandra_comparison} we show the energy-independent Chandrasekhar's single-scattering results for reflection \citep{Chandrasekhar1960} for the same three cases of incident polarization and $M_\mathrm{i}$ values, displayed in the same manner as in Figure \ref{disc_mue_dependent}. We numerically integrated the analytical predictions in all azimuthal emission angles $\Phi_\textrm{e}$ and selected ranges of incident angles. The analytical reflection model matches in the total polarization degree and direction predictions by {\tt xsstokes\_disc} rather in the lower energy band studied for any $M_\mathrm{i}$, as there the {\tt xsstokes\_disc} results are absorption dominated; and at nearly all X-ray energies for low $M_\mathrm{i}$, where the {\tt xsstokes\_disc} code provides rather energy-independent results.

\section{Conclusions}\label{conclusions}

We developed a series of new routines that help to predict X-ray polarization arising from reflection on distant axially symmetric scatterers illuminated by a central isotropic power-law emission of arbitrary polarization. A simple non-relativistic integration across the inner walls of a circular torus, using local reflection tables originally computed for coronal power-law impinging on constant-density AGN accretion disc, can in some configurations well represent the expected outcome from reprocessing in cold AGN dusty tori. 
This was tested against MC simulations that in addition to our new simple model allow for partial transparency, changeable ionization and self-irradiation effects. Using a much simpler approach, we obtained in the most similar cases still about twice higher net polarization fraction (tens of \%), if the dominant resulting polarization direction was perpendicular to the axis of symmetry, and about twice lower (a few \%), if the net polarization direction was parallel. Detailed investigation shows that the main driver of the difference in the model predictions is the partial transparency related to the absorber's structure. Contribution from the inner walls alone to the total AGN emission is difficult to estimate when decomposing observational signal in type-2 AGNs and leads to increase in perpendicularly polarized X-ray contribution. Searching for the least similar case, we showed that the MC simulations for optically thin covering media reach lower tens of \% (reprocessed) polarization with parallel orientation, which is rather independent of the opening angle, but dependent on inclination.

The difference in predicted polarization between the two extreme cases (the optically thin MC case and the new model representing reflection from optically thick inner walls) is generally smaller for high half-opening angles and low inclination of the observer, when the partial transparency plays less role. Our analysis of certain regimes in the MC code shows that it is possible to span continuously through all values of polarization in between the two extremes by varying the torus density and ionization. Albeit many of the intermediate results presented are only illustrative to explain the role of different model parameters, the broad analysis also cross-validates the implementation of relative size and position parameters of the equatorial scatterer in the new model. Computation of more realistic models with non-homogeneous ionization profiles of all species and the focus on spectral lines, non-stationarity or geometries other than a torus is beyond the scope of this work and cannot be examined briefly due to large number of parameters.

Nonetheless, the routines developed for these estimates can be already of wider use. The torus integrating routine is wavelength independent and the calculations can be easily modified to elliptical torus profiles. The local reflection tables from \cite{Podgorny2021} used for the distant X-ray reprocessing discussed in this paper can be also easily replaced by any other reflection tables in a similar FITS format. For example, by ionization-dependent tables, or the same nearly neutral computations but for different densities, or analytical prescriptions, as already briefly shown. Therefore, we suggest a possible usefulness of the torus integration routine for studying physical objects also other than dusty AGN tori, such as thick accretion discs or obscuring outflows of accreting stellar-mass black holes, neutron stars or white dwarfs.

One could also separately precompute the resulting FITS spectropolarimetric tables for arbitrary axially symmetric reflector for three independent incident polarization states, using any method and e.g. the ASCII to FITS convertor that we applied. Employing new model parameters, it is then easy to adapt the follow-up {\tt xsstokes} interpolation routine, which represents a fast {\tt XSPEC} compatible fitting model interpolating for arbitrary incident polarization state. In order to test such possibility, we provided another example of a simple non-relativistic distant disc reflector of a central power-law originating in arbitrarily extended inner coronal region. The results for small extension of the X-ray source are consistent with the centrally illuminated reflecting toroidal structures studied for large half-opening angles and with other X-ray polarization studies of reflection from accretion discs, assuming compact coronal geometries.

\section*{Acknowledgements}

All authors thank Michal Bursa, who created the documentation, the basis and final form of the {\tt XSPEC Table Model Generator} routine, which JP only checked for consistency with a similar routine independently developed. The authors acknowledge thorough reading of the manuscript by the referee and very helpful comments leading to substantial improvement of this work. The authors thank also Giorgio Matt for useful discussions in the early stages of the work and the Strasbourg Astronomical Observatory for providing the necessary computational capacities. JP and MD acknowledge the support from the Czech Science Foundation project GACR 21-06825X and the institutional support from the Astronomical Institute RVO:67985815. The background for Figure \ref{xsstokes_mo} was obtained with the {\tt GeoGebra} tool available at \url{https://www.geogebra.org/m/kxwUhFqx}.

\section*{Data Availability}

The {\tt XSPEC} software is publicly available in the HEASARC database (\url{https://heasarc.gsfc.nasa.gov/xanadu/xspec/}). The {\tt xsstokes\_torus} and {\tt xsstokes\_disc} C routines for {\tt XSPEC} and their source tables and documentation are available at \url{https://github.com/jpodgorny/xsstokes_torus} and \url{https://github.com/jpodgorny/xsstokes_disc}, respectively. The Python 3 routine {\tt torus\_intergrator} and its documentation is available at \url{https://github.com/jpodgorny/torus_integrator}. The Python 3 routine {\tt XSPEC Table Model Generator} and its documentation is available at \url{https://github.com/mbursa/xspec-table-models}. The {\tt STOKES} code version \textit{v2.07} is currently not publicly available and it will be shared upon reasonable request.



\bibliographystyle{mnras}
\bibliography{example} 




\appendix

\section{Numerical implementation of the torus reflection}\label{technical_details}

In this section, we provide details on the computational routines that lead to the physical model of reflection on a geometrical torus, described in Section \ref{torus_reflection}. The parametrization is given in Figure \ref{xsstokes_mo}.

We first use a simple integrator of local reflection tables across the inner walls of the torus. The local reflection tables \citep{Podgorny2021} store numerically computed Stokes parameters $I$, $Q$ and $U$ dependent on energy (in 300 logarithmically spaced energy bins between 0.1 and 100 keV), incident power-law index $\Gamma$, local incident and emission cosines of inclination angles $\mu_\textrm{i}$ and $\mu_\textrm{e}$, and the azimuthal emission angle $\Phi_\textrm{e}$ -- all for three independent polarization states of the incident radiation: unpolarized and $100\%$ polarized perpendicularly and with a $45\degr$ offset from the local surface's projected normal (see below). The Python 3 module, called {\tt torus\_integrator}, that computes the total reflection output in ASCII format is available in the Data Availability section alongside user instructions and documentation. The detailed calculations are described in Appendix \ref{torus_calculus}. Inside this routine, which is not yet the {\tt xsstokes\_torus} model for {\tt XSPEC}, the user can define an isotropically emitting power-law point source located at the center of the coordinates with $1.2 \leq \Gamma \leq 3.0$ and a state of incident polarization given by the primary polarization degree $p_0$ and primary polarization direction $\Psi_0$. These model parameters should follow the values given by the local reflection tables, as the interpolation is not performed for these parameters. Then the user sets any torus half-opening angle $25\degr \leq \Theta < 90\degr$ measured from the pole and any observer's inclination $0\degr < i < 90\degr$ measured from the pole. In the current setup, the results are independent of the distance of the torus from the center, i.e. the torus inner radius $r_\textrm{in}$.

There are two options via a parameter $B$, whether the integrating routine takes into account the entire observable and reflecting torus surface ($B = 1$), or only the illumination and observation of the upper half-plane of the torus ($B = 0$, expecting some optically thick absorbing material extending from the outer accretion disc to $r_\textrm{in}$). The computational grid for local reflection that is interpolated linearly is given in the angular coordinates $u \in (0, 2\pi]$ and $v \in [\pi - \Theta, \pi + \Theta]$ (for $B = 1$) or $v \in [\pi - \Theta, \pi]$ (for $B = 0$) and by linear binning defined by user via $2N_\textrm{u}$ and $N_\textrm{v}$ points in the corresponding $u$ and $v$ ranges. We have a good experience with convergence to an exact result for $N_\textrm{u} > 30$ and $N_\textrm{v} > 60$ in the most computationally demanding corners of the parameter space. We provide access in the Data Availability section to precomputed full tables for {\tt xsstokes\_torus} with $N_\textrm{u} = 50$ and $N_\textrm{v} = 80$, which were used for the results presented in this paper. The energy resolution of the output remains from the local reflection tables and the output is stored in ASCII file format with lower and upper energy bin edges and the Stokes $I$, $Q$ and $U$ columns. The {\tt torus\_integrator} has also a built-in imaging function, so that the user can have more insight into the spatially integrated results.

In the next step, the ASCII files are converted into FITS files conforming to the OGIP standard of the {\tt XSPEC} version \textit{12.13.0} suitable for polarization, because only in this format the subsequent {\tt xsstokes\_torus} model reads them. In the Data Availability section we provide the Python 3 conversion script called {\tt XSPEC Table Model Generator} that was used for the conversion. A reduction of the code to a pure spectral FITS file useful for previous {\tt XSPEC} versions is straight forward within.

\begin{figure}
    \centering
    \includegraphics[width=\columnwidth]{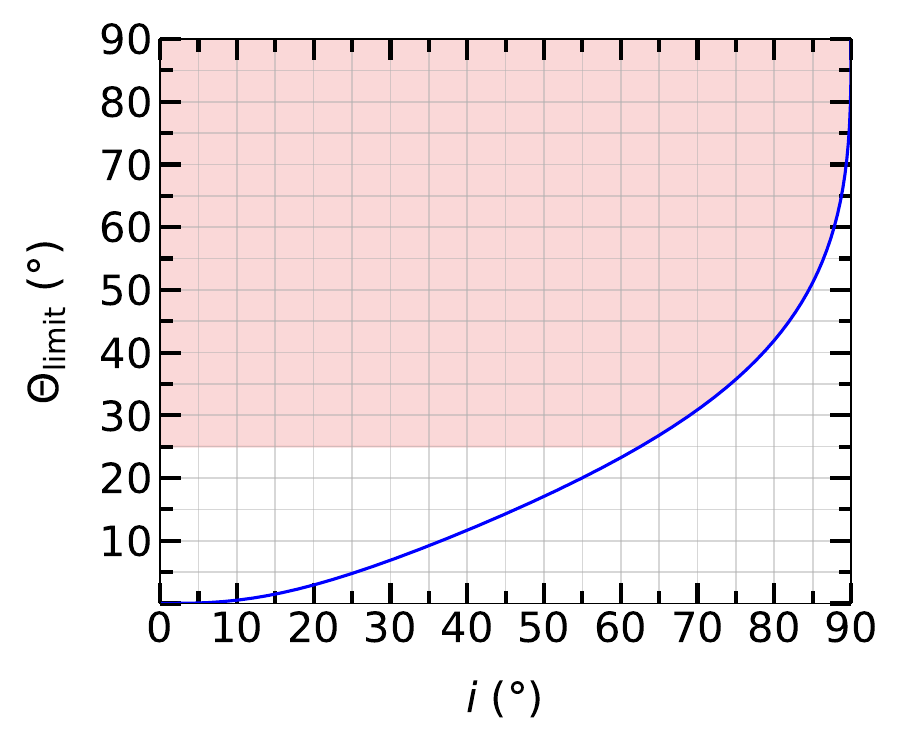}
    \caption{The blue curve is the limiting half-opening angle $\Theta_\mathrm{limit}$ with inclination $i$, above which the reflecting area of the torus is directly observable. Such curve naturally appears for the torus geometry with a convex inner boundary with respect to the central source. Any half-opening angle below this curve for a given $i$ leads to a full eclipse. The computation is described in Appendix \ref{torus_calculus}. The shaded area represents the range of $\Theta(i)$ between $\Theta_\textrm{min}(i)$ and $\Theta_\textrm{max}(i)$, where all presented codes smoothly operate and where we define $\Theta'(\Theta)$.}
    \label{terminator_function}
\end{figure}

Finally, we provide in the Data Availability section the {\tt xsstokes\_torus} C routine compatible with {\tt XSPEC}, including the necessary FITS table dependencies obtained in the previous step. Table \ref{xsstokes_options} (top) introduces the available parameters. Note that in this final {\tt XSPEC} model, we have to rescale the half-opening angle due to the no-transparency condition. When the illuminated surface is fully obscured (see Appendix \ref{torus_calculus}), the {\tt torus\_integrator} stores zeros for the resulting Stokes parameters, which can cause interpolation issues in {\tt XSPEC}. Therefore, we introduce a transformation
\begin{equation}
    \Theta' = \frac{\Theta-\Theta_\textrm{min}(i)}{\Theta_\textrm{max}-\Theta_\textrm{min}(i)}
\end{equation}
that allows the new interpolation parameter $\Theta'$ to vary between 0 and 1, effectively spanning the allowed values of $\Theta$ that depend on $i$. The $\Theta_\textrm{max}$ value is chosen as $90\degr$ and the $\Theta_\textrm{min}(i)$ value is chosen as $\max\{25\degr,\Theta_\textrm{limit}(i)\}$. The theoretical $\Theta_\textrm{limit}(i)$ curve for the torus terminator is shown on Figure \ref{terminator_function} and is numerically precomputed and stored in a text file. The {\tt xsstokes\_torus} C routine provides automatic conversion back to $\Theta$ for the model output provided, if used outside of {\tt XSPEC}.

Thanks to the axial symmetry of our toroidal setup and the neglecting of relativistic effects, the incident polarization direction does not change with respect to the local reflection frame for any point on the torus surface. The reflected Stokes vector $\Vec{S} = (I, Q, U)$ (we assume Stokes $V = 0$) for arbitrary primary polarization state can be decomposed into a basis of Stokes vectors computed for three independent primary polarization states \citep[e.g.][]{Chandrasekhar1960}. Because we have one and the same $p_0$ and $\Psi_0$ for all reflecting surface points and because of the linearity of Stokes parameters, we can precompute the \textit{total} integrated Stokes vector $\Vec{S}_\mathrm{tot}$ for only three independent incident states of polarization, each stored in a FITS file, and compose the result for arbitrary incident polarization. We have chosen to store the computations with {\tt torus\_integrator} and {\tt XSPEC Table Model Generator} for unpolarized and $100\%$ perpendicularly ($\Psi_0 = \pi/2$) and $100\%$ diagonally ($\Psi_0 = \pi/4$) polarized radiation. In this way we are able to use equation (2) from \cite{Podgorny2023} in {\tt xsstokes\_torus} in the same form. It allows to effortlessly interpolate arbitrary incident polarization in {\tt XSPEC} for any global $p_0$ and $\Psi_0$ parameters. In the same way, the core of the {\tt xsstokes\_torus} routine can serve for any axially symmetric reflecting model with $\Vec{S}$ produced for three independent polarization states: $\Vec{S}_\textrm{tot}(p_0 = 0, \Psi_0 = \textrm{undefined})$, $\Vec{S}_\textrm{tot}(p_0 = 1, \Psi_0 = \pi/2)$, and $\Vec{S}_\textrm{tot}(p_0 = 1, \Psi_0 = \pi/4)$ for different parameters (related to the global geometry and local reprocessing). See Appendix \ref{torus_calculus} on how each component of $\Vec{S}_\textrm{tot}$ is obtained for the reflecting torus geometry to give
\begin{equation}\label{stot}
\begin{split}
    \Vec{S}_\textrm{tot}(p_0 & ,\Psi_0) = \, \Vec{S}_\textrm{tot}(0, -) \, 
    \\
    &\quad  + \, p_0\{ [\Vec{S}_\textrm{tot}(0, -) - \, \Vec{S}_\textrm{tot}(1, \pi/2)] \cos{2\Psi_0 }
    \\
    &\quad + [\Vec{S}_\textrm{tot}(1, \pi/4) - \Vec{S}_\textrm{tot}(0, -)]\sin{2\Psi_0} \} \textrm{ .}
\end{split}
\end{equation}
We additionally introduce $\Delta\Psi$ as the orientation of the system with respect to the global axis of symmetry, which is also a free parameter in {\tt xsstokes\_torus} and to which the output Stokes $Q_\textrm{tot}$ and $U_\textrm{tot}$ parameters can be rotated as a last step.
\begin{table*}
\begin{center}
\caption{The {\tt XSPEC} parameters available in the two attached {\tt xsstokes} model versions.}
\resizebox{2.\columnwidth}{!}{
\begin{tabular}{p{2cm}p{2cm}p{4.5cm}p{1cm}p{1cm}p{1cm}p{1cm}p{1cm}p{0.8cm}p{1.5cm}p{0.5cm}}
\hline
\hline
model name & Parameter (unit) & Description & Initial & Min & Bottom & Max & Top & Step & Interpolation & Free  \\  
\hline
\hline
{\tt xsstokes\_torus} & $\Gamma$ & primary power-law photon index & 2.0 & 1.2 & 1.2 & 3.0 & 3.0 & 0.1 & linear & yes \\ 
 & $\cos{(i)}$ & cosine of observer's inclination & 0.775 & 0.025 & 0.025 & 0.975 & 0.975 & 0.01 & linear & yes \\ 
 & $\Theta'$ & rescaled half-opening angle & 0.5 & 0. & 0. & 1. &  1. & 0.05 & linear & yes \\ 
  & $B$ &  integrating below the equator yes/no &  1. &  0. &  0. &  1. &  1. &  0.1 & linear & no \\ 
 & $p_0$ & primary polarization degree & 0. & 0. & 0. & 1. & 1. & 0.01 & linear & yes \\ 
 & $\Psi_0$ (\degr) & primary polarization direction & 0. & -90. & -90. & 90. & 90. & 5 & linear & yes \\ 
  & $\Delta\Psi$ (\degr) & system orientation & 0. & -90. & -90. & 90. & 90. & 5 & linear & no \\ 
 & $z$ & overall Doppler shift & 0 & -0.999 & -0.999 & 10 & 10 & 0.1 & linear & no \\ 
 & Stokes & output definition (see documentation) & 1 & -1 & -1 & 10 & 10 & 1 & linear & no \\ 
\hline
{\tt xsstokes\_disc} 
 & $M_\mathrm{i}$ & coronal size scaling & 0.3 & 0.2 & 0.2 & 1.0 & 1.0 & 0.01 & linear & yes \\ & $\Gamma$ & primary power-law photon index & 2.0 & 1.2 & 1.2 & 3 & 3 & 0.1 & linear & yes \\ 
 & $\cos{(i)}$ & cosine of observer's inclination & 0.775 & 0.025 & 0.025 & 0.975 & 0.975 & 0.01 & linear & yes \\ 
 & $p_0$ & primary polarization degree & 0. & 0. & 0. & 1. & 1. & 0.01 & linear & yes \\ 
 & $\Psi_0$ (\degr) & primary polarization direction & 0. & -90. & -90. & 90. & 90. & 5 & linear & yes \\ 
  & $\Delta\Psi$ (\degr) & system orientation & 0. & -90. & -90. & 90. & 90. & 5 & linear & no \\ 
 & $z$ & overall Doppler shift & 0 & -0.999 & -0.999 & 10 & 10 & 0.1 & linear & no \\ 
 & Stokes & output definition (see documentation) & 1 & -1 & -1 & 10 & 10 & 1 & linear & no \\
\hline
\end{tabular}
}
\label{xsstokes_options}
\end{center}
\end{table*}

\section{Toroidal integrating routine: detailed description}\label{torus_calculus}

We will provide a few computational details of the routine {\tt torus\_integrator} that is introduced in Appendix \ref{technical_details}. The code uses standard relations between the cartesian coordinate system \{$x$, $y$, $z$\} given by the base vectors $\Vec{e}_\textrm{x} = (1,0,0)$, $\Vec{e}_\textrm{y} = (0,1,0)$, $\Vec{e}_\textrm{z} = (0,0,1)$ and the torus surface coordinates \{$u$, $v$\} given by the base vectors $\frac{\partial}{\partial u} = (-\sin{u}, \cos{u}, 0)$, $\frac{\partial}{\partial v} = (-\sin{v}\cos{u}, -\sin{v}\sin{u}, \cos{v})$:
\begin{equation}\label{coordtrafo}
	\begin{aligned}
		x &= (R+r\cos{v})\cos{u} \\
		y &= (R+r\cos{v})\sin{u} \\
            z &= r\sin{v} \textrm{ ,}
	\end{aligned}
\end{equation}
where $R = r + r_\textrm{in}$ is the distance between (0,0,0) and the center of the toroidal circle, and
\begin{equation}
    r = r_\textrm{in}\frac{\cos{\Theta}}{1-\cos{\Theta}}
\end{equation}
is the radius of the toroidal circles in the meridional plane (see Figure \ref{xsstokes_mo}). The surface is given by another standard implicit formula
\begin{equation}
\Phi = (x^2 + y^2 + z^2 + R^2 - r^2)^2 - 4R^2(x^2+y^2) = 0 \textrm{ .}
\end{equation}

\begin{figure}
	\includegraphics[width=0.58\columnwidth]{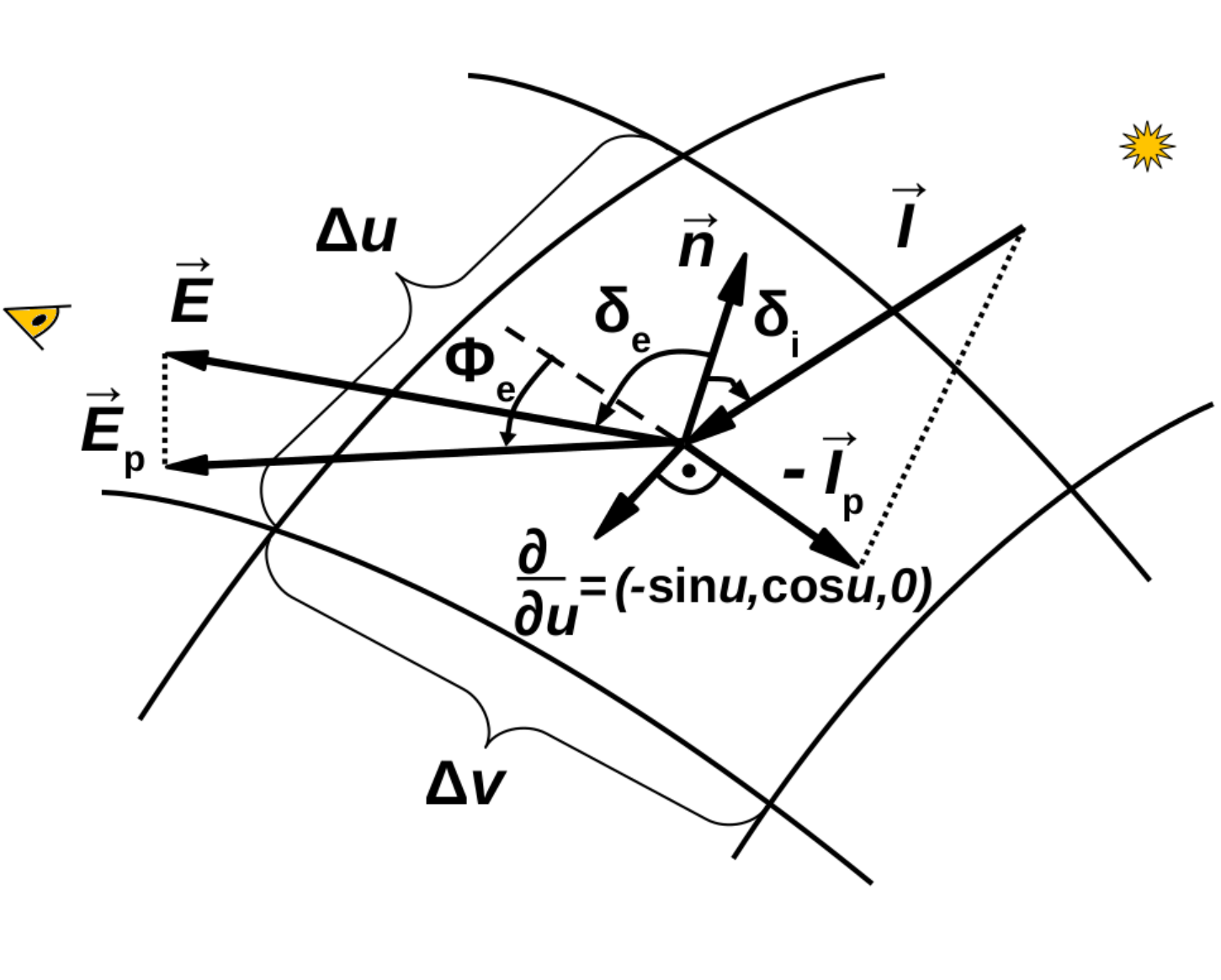}
        \includegraphics[width=0.4\columnwidth]{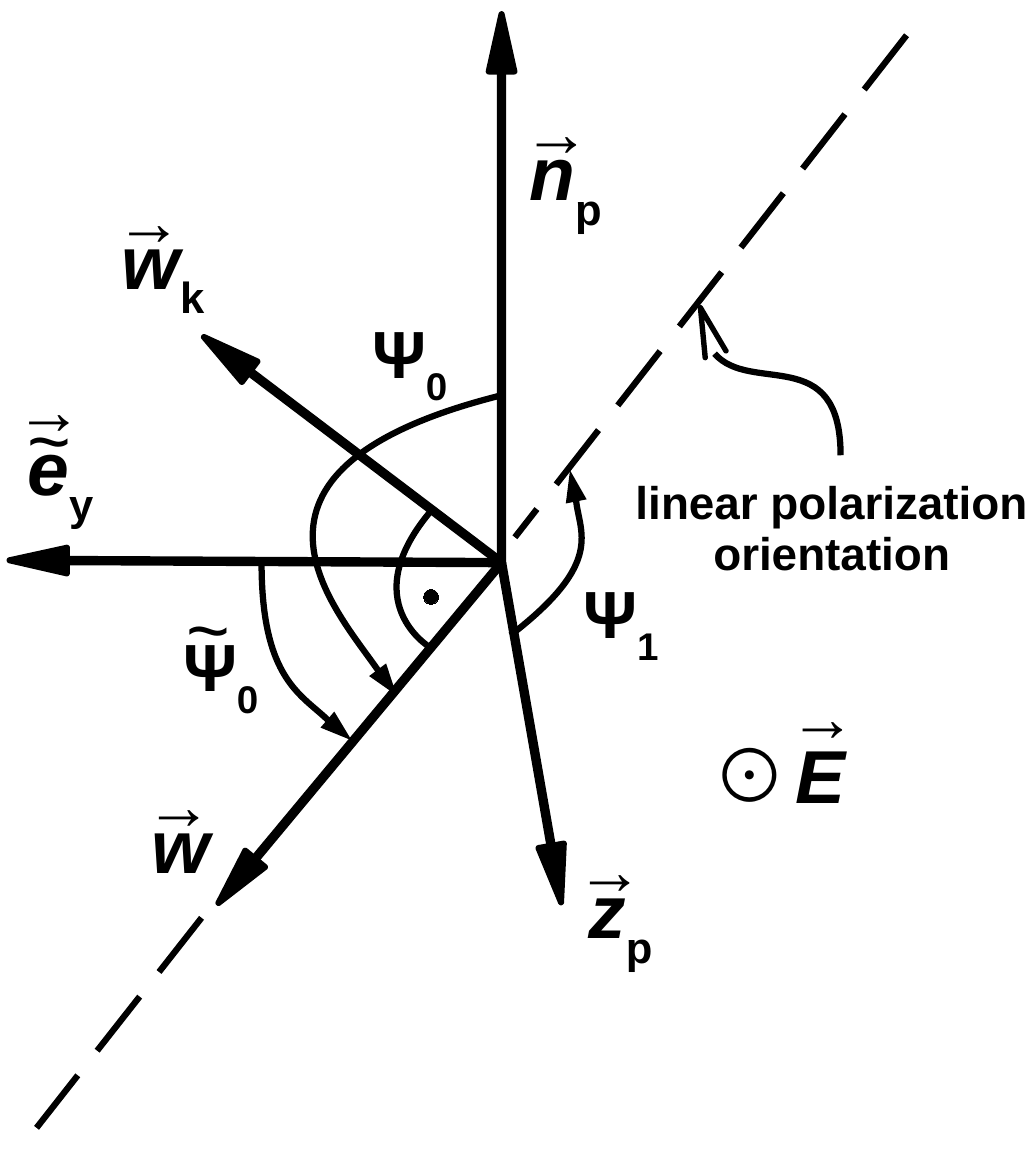}
	\caption{Left: the definitions of angles and vectors in the locally reflecting plane, tangent to the torus surface. Right: the definitions of angles and vectors in the polarization plane, perpendicular to the photon momentum direction. These are needed for the calculations of rotation of the polarization direction by changing frame of reference.}
	\label{local_angles}
\end{figure}

The local reflection tables need to be interpolated at each point in the \{$u$, $v$\} grid in their three angular dependencies $\mu_\textrm{i} = \cos{\delta_\textrm{i}}$, $\mu_\textrm{e} = \cos{\delta_\textrm{e}}$, and $\Phi_\textrm{e}$ properly defined in \cite{Dovciak2011} and depicted here in Figure \ref{local_angles} (left). The code precomputes these angles at each grid point given their definitions in the local frame and given the local surface normal $\Vec{n} = \frac{\nabla \Phi}{\lvert \nabla \Phi \rvert}$, the incident vector $\Vec{I} = (x, y, z)$, the emission vector $\Vec{E} = (0, \sin i, \cos i)$, and their respective projections to the local tangent plane through scalar products.

The code ensures that the grid points with $\delta_\textrm{i} \notin [0, \frac{\pi}{2}]$ and $\delta_\textrm{e} \notin [0, \frac{\pi}{2}]$ are not accounted for. For $B = 0$ we limit $\pi - \Theta \leq v \leq \pi$, i.e. we do not account for those regions that lie below the equatorial plane where some outer disc material could be extending, and for any $v$ smaller than the shadow boundary given by $\Theta$. For $B = 1$ we limit $\pi - \Theta \leq v \leq \pi + \Theta$, symmetrically with respect to the equatorial plane. The region of integration has to be also visible by the observer, which is for each $u$ and $i$ given by the limiting curve of no-transparency condition $0 = \Vec{n} \cdot \Vec{E}$. Therefore, after a few simple steps we obtain a prescription for this curve
\begin{equation}
    v_\textrm{limit}(u,i) = -\arctan ( 2 \sin u \tan i) + \pi \textrm{ .} 
\end{equation}
Lastly, we also have to account for self-obscuration of the reflected rays by the opposite side of the torus, closer to the observer (the scenario for highly inclined observers). To find the torus terminator, we define a grazing vector $\Vec{G}$ that begins at $\Vec{X}_1=$ ($x_1$, $y_1$, $z_1$) at the opposite half-space from the observer given by $\pi \leq u \leq 2\pi$ and some $v_\textrm{self-obs}$ per each $u$. The grazing vector points in the $\Vec{E}$ direction towards the observer and ends at a point $\Vec{X}_2=$ ($x_2$, $y_2$, $z_2$) tangential to the torus surface, i.e. defined by $v_\textrm{limit}$ for unknown $0 \leq u_\textrm{t} \leq \pi$. Thus, parametrically for $\Vec{X}_1 + t\Vec{E} = \Vec{X}_2$ we write
\begin{equation}\label{cond1}
	\begin{aligned}
		(R + r\cos v_\textrm{self-obs})\cos u + tE_1 &= (R + r\cos v_\textrm{limit})\cos u_\textrm{t} \\
		(R + r\cos v_\textrm{self-obs})\sin u + tE_2  &=  (R + r\cos v_\textrm{limit})\sin u_\textrm{t} \\
            r\sin v_\textrm{self-obs} + tE_3 &=  r\sin v_\textrm{limit} \textrm{ .}
	\end{aligned}
\end{equation}
Therefore, $v_\textrm{self-obs}(u, i)$ is found when we search for a solution \{$v_\textrm{self-obs}, t, u_\textrm{t}$\} of a set of equations
\begin{equation}\label{cond2}
	\begin{aligned}
		0 &= (R + r\cos v_\textrm{self-obs})\cos u - (R + r\cos v_\textrm{limit}(u_\textrm{t},i) )\cos u_\textrm{t} \\
		0 &= (R + r\cos v_\textrm{self-obs})\sin u + t\sin i - (R + r\cos v_\textrm{limit}(u_\textrm{t},i) )\sin u_\textrm{t} \\
            0 &= r\sin v_\textrm{self-obs} + t\cos i - r\sin v_\textrm{limit}(u_\textrm{t},i) \textrm{ .}
	\end{aligned}
\end{equation}

The acquisition of the curve shown in Figure \ref{terminator_function} is straightforward from here. When increasing the half-opening angle $\Theta$ from 0 for a given $i$, the first observable point from the reflecting area will be located at $u = 3\pi/2 = u_\mathrm{t} + \pi$ (in the direction away from the observer) at $v_\mathrm{self-obs}(3\pi/2,i) = \pi - \Theta_\mathrm{limit}$ (when crossing the upper shadow boundary). The set of Equations (\ref{cond2}) then reduces to
\begin{equation}
\begin{split}
    0 &= \cos \Theta_\mathrm{limit} + \tan i (\sin v_\mathrm{limit}(\pi/2,i) - \sin \Theta_\mathrm{limit}) \\ & \quad \quad \quad -\frac{2}{\cos \Theta_\mathrm{limit}} - \cos v_\mathrm{limit}(\pi/2,i) \textrm{ ,}
\end{split}
\end{equation}
where
\begin{equation}
    v_\mathrm{limit}(\pi/2,i) = -\arctan ( 2 \tan i) + \pi \textrm{ .}
\end{equation}
The function $\Theta_\mathrm{limit}(i)$ can be evaluated numerically from here and tabulated.

Each local scattering surface in the user-defined grid is given by boundaries $v_1(u,v)$, $v_2(u,v)$, $u_1(u,v)$, $u_2(u,v)$ that are defined in linear binning via $N_\textrm{v}$ and $2N_\textrm{u}$ points (the number of bins in $u$ covers only the $[\pi/2,3\pi/2]$ interval, because the other half is symmetrically added). Or the local surface is set to zero, if the central bin point \{$u$, $v$\} does not fall within the restricting boundary conditions above, which is a good approximation for sufficiently high resolution. We then approximate the contributing local scattering surfaces by a tangent rectangle and compute its area:
\begin{equation}
\begin{split}
    A_\textrm{u,v} & = \int_{u_1}^{u_2} \int_{v_1}^{v_2} r(R+r\cos v) dv du \\
    & \approx r(R+r\cos v)(v_2-v_1)(u_2-u_1) \textrm{ .}
\end{split}
\end{equation}
At each \{$u,v$\} we perform a tri-linear interpolation in $\mu_\textrm{i}$, $\mu_\textrm{e}$, and $\Phi_\textrm{e}$ of the tabular Stokes parameters $I$, $Q$ and $U$, commonly denoted as $\Vec{S}(\mu_\textrm{i},\mu_\textrm{e},\Phi_\textrm{e}; E, \Gamma, p_0, \Psi_0)$, in order to obtain $\Vec{\Bar{S}}_\textrm{u,v}(\mu_\textrm{i}(u,v),\mu_\textrm{e}(u,v),\Phi_\textrm{e}(u,v); E, \Gamma, p_0, \Psi_0)$. The Stokes parameters stored in the tables from \cite{Podgorny2021} also depend on energy $E$, power-law index $\Gamma$ and primary polarization state ($p_0, \Psi_0$), but in the current version we require the exact stored values to be requested for the final ASCII files in {\tt torus\_integrator}, in order not to interpolate further in these for computational efficiency.

The polarization vector of the interpolated reflected rays is in addition rotated from the local frame to the global frame. In global coordinates, the Stokes parameters are set with respect to the torus axis of symmetry. The incident electric vector orientation does not have to be rotated to conform to the polarization definition in the local frame due to the axial symmetry of the entire torus surface and emission placed in the central point (see Appendix \ref{technical_details}). It holds that the polarization fraction after the rotation $p_1$ is equal to the $p_0$ in the local reflection frame, and similarly for the intensity: $I_1 = I_0$. Thus, for the remaining Stokes parameters we compute $Q_1 = \sqrt{Q_0^2 + U_0^2}\cos{(2\Psi_1)}$ and $U_1 = \sqrt{Q_0^2 + U_0^2}\sin{(2\Psi_1)}$. In order to obtain the $\Psi_1(\Psi_0)$ function we will use Definition (\ref{ppsidef}). See Figure \ref{local_angles} (right) for a schematic drawing of the angles and vectors in the polarization plane. We define $\Psi_0$ with respect to the projected normal $\Vec{n}_\textrm{p} = \Vec{n} - (\Vec{n} \cdot \Vec{E})\Vec{E}$ and $\Psi_1$ with respect to the projected global axis of symmetry $\Vec{z}_\textrm{p} = \Vec{e}_\textrm{z} - (\Vec{e}_\textrm{z} \cdot \Vec{E})\Vec{E}$. Then $\Vec{\Tilde{e}}_\textrm{y} = \Vec{E} \times \frac{\Vec{n}_\textrm{p}}{\lvert \Vec{n}_\textrm{p} \rvert}$, $\Vec{\Tilde{e}}_\textrm{x} = \frac{\Vec{n}_\textrm{p}}{\lvert \Vec{n}_\textrm{p} \rvert}$ and $\Vec{\Tilde{e}}_\textrm{z} = \Vec{E}$ are the base vectors in the polarization plane and the polarization vector is locally given by $\Vec{w}_\textrm{loc} = (\cos{\Psi_0}, \sin{\Psi_0}, 0)$. The conditions $\cos{\Psi_0} = \Vec{\Tilde{e}}_\textrm{x} \cdot \Vec{w}$, $\cos{\Tilde{\Psi}_0} = \sin{\Psi_0} = \Vec{\Tilde{e}}_\textrm{y} \cdot \Vec{w}$ and $\Vec{w} \cdot \Vec{E} = 0$ give
\begin{gather}
 \Vec{w}_\textrm{loc}
 = 
  \begin{pmatrix}
   \Vec{\Tilde{e}}_\textrm{x}, &
   \Vec{\Tilde{e}}_\textrm{y}, & \Vec{\Tilde{e}}_\textrm{z}
   \end{pmatrix} \cdot \Vec{w} \equiv \beta \cdot \Vec{w}.
\end{gather}
Hence, we need to invert this matrix equation to obtain $\Vec{w} = \beta^{-1} \cdot \Vec{w}_\textrm{loc}$, in order to also calculate $\Vec{w}_\textrm{k} = - \Vec{E} \times \Vec{w}$. Then
\begin{equation}
    \Psi_1 = 
	\begin{cases}
		\arccos{ \left( \dfrac{\Vec{z}_\textrm{p} \cdot \Vec{w}}{\lvert \Vec{z}_\textrm{p} \rvert \lvert \Vec{w} \rvert} \right) } \ , & \textrm{if} \ \arccos{\left( \dfrac{\Vec{z}_\textrm{p} \cdot \Vec{w}_\textrm{k}}{\lvert \Vec{z}_\textrm{p} \rvert \lvert \Vec{w}_\textrm{k} \rvert} \right)} \leq \dfrac{\pi}{2} \ \\
		-\arccos{ \left( \dfrac{\Vec{z}_\textrm{p} \cdot \Vec{w}}{\lvert \Vec{z}_\textrm{p} \rvert \lvert \Vec{w} \rvert} \right) }  \ , & \textrm{if} \ \arccos{\left( \dfrac{\Vec{z}_\textrm{p} \cdot \Vec{w}_\textrm{k}}{\lvert \Vec{z}_\textrm{p} \rvert \lvert \Vec{w}_\textrm{k} \rvert} \right)} > \dfrac{\pi}{2} \ .
	\end{cases}
\end{equation}

We finish by numerically integrating the Stokes parameters from locally tangent planes projected to the line of sight across all \{$u,v$\} that fulfill the aforementioned shadow and visibility conditions:
\begin{equation}
    \Vec{S}_\textrm{tot}(i, \Theta, p_0, \Psi_0, \Gamma, E) = K \sum_\textrm{u,v} A_\textrm{u,v} \mu_\textrm{e}(u,v)\mu_\textrm{i}(u,v) \frac{1}{\varrho(u,v)^{2}} \Vec{\Bar{S}}_\textrm{u,v} \textrm{ ,}
\end{equation}
with weighting according to the locally illuminating flux dependent on distance from the center $\varrho(u,v)$ as $\sim \frac{1}{\varrho(u,v)^{2}}$. Because of the weighting with $\varrho(u,v)$ that scales linearly with $r_\textrm{in}$, the results are independent of $r_\textrm{in}$. In our approximation, we take for all surface points the most neutral version of the local reflection tables and the ionization parameter is not dependent on the local flux received. The output $\Vec{S}_\textrm{tot}$ is renormalized via $K$ constant for storage convenience. We save the results in a user-defined directory in the ASCII file format. Such $\Vec{S}_\textrm{tot}$ tables, if converted to FITS, are then ready to be used inside {\tt xsstokes\_torus} for any incident polarization, following Equation (\ref{stot}).

\section{Numerical implementation of the disc reflection}\label{technical_disc}

In this section, we provide details on the computations that lead to the physical model of reflection on a distant accretion disc, described in Section \ref{disc_reflection}. We first created an ASCII version of the most neutral part of the disc reflection tables from \cite{Podgorny2021} that store the original tables uniformly added in all $\Phi_\textrm{e}$ and $0 \leq \mu_\textrm{i} = \cos{\delta_\textrm{i}} \leq M_\mathrm{i}$ for three independent incident polarization states and nine linearly spaced values of $M_\mathrm{i} \in [0.2;1]$. Then they were converted to FITS using the {\tt XSPEC Table Model Generator}. We preserved the energy binning and dependency on $\mu_\textrm{e} = \cos(i)$ and $\Gamma$. These constitute the {\tt XSPEC} fitting parameters in the follow-up {\tt xsstokes\_disc} C routine, alongside polarization. This routine operates on the same basis as the {\tt xsstokes\_torus} routine and requires the aforementioned FITS tables as dependencies. It interpolates for arbitrary incident polarization $p_0$ and $\Psi_0$ through Equation (\ref{stot}) from three precomputed tables in independent incident polarization states. It is possible to adjust the final Stokes parameters $Q_\mathrm{tot}$ and $U_\mathrm{tot}$ by setting a global system orientation $\Delta\Psi$ with respect to the disc normal. Table \ref{xsstokes_options} (bottom) summarizes the resulting model parameters available for data fitting at once.

\section{The effect of changing primary polarization}\label{incident_polarization}

Theoretical models of X-ray reflection in accreting systems are often assuming unpolarized source of irradiation, which is not supported by theoretical studies of Comptonization in coronae \citep[e.g.][]{Poutanen2018, Krawczynski2022}, nor by the recent \textit{IXPE} observations \citep{Krawczynski2022b, Gianolli2023, Veledina2023b}. Thus, here we provide an overview of the effect of changing incident polarization in axially symmetric reflectors, where we can use Equation (\ref{stot}) (or any other variant of it for different basis) that decomposes the signal into the basis of three independent incident polarization states. Such investigations are useful for predictions of the change of resulting polarization in an existing reflection model that uses unpolarized irradiation, given that the incident polarization changes by $p_0$.

Let us denote the total (integrated) Stokes vector for particular incident polarization states as $\Vec{S}_\textrm{tot}(0,-) \equiv \Vec{S}_\textrm{U}$, $\Vec{S}_\textrm{tot}(1, \pi/2) \equiv \Vec{S}_\perp$, and $\Vec{S}_\textrm{tot}(1, \pi/4) \equiv \Vec{S}_\times$. Then combining Equations (\ref{ppsidef}) and (\ref{stot}) we have for the resulting total linear polarization degree:
\begin{equation}
\begin{split}
    p & = \dfrac{\sqrt{Q_\mathrm{tot}^2+U_\mathrm{tot}^2}}{I_\mathrm{tot}} \\
    & = \left\{ \left\{\dfrac{ Q_\mathrm{U}+p_0[(Q_\mathrm{U} - Q_\perp)\cos{2\Psi_0 } + (Q_\times - Q_\mathrm{U})\sin{2\Psi_0}] }{I_\mathrm{U}+p_0[(I_\mathrm{U} - I_\perp)\cos{2\Psi_0 } + (I_\times - I_\mathrm{U})\sin{2\Psi_0} ]} \right\}^2 \right.\\
    & \left. + \left\{\dfrac{ U_\mathrm{U}+p_0[(U_\mathrm{U} - U_\perp)\cos{2\Psi_0 } + (U_\times - U_\mathrm{U})\sin{2\Psi_0}] }{I_\mathrm{U}+p_0[(I_\mathrm{U} - I_\perp)\cos{2\Psi_0 } + (I_\times - I_\mathrm{U})\sin{2\Psi_0} ]} \right\}^2 \right\}^{\frac{1}{2}} \textrm{ .}
\end{split}
\end{equation}
Assuming that $p_0$ is small to the linear order, this reduces to
\begin{equation}
\begin{split}
    p & = \dfrac{\sqrt{Q_\mathrm{U}^2 + U_\mathrm{U}^2}}{I_\mathrm{U}} + p_0 \left[ \cos{2\Psi_0 } \left( \dfrac{I_\perp \sqrt{Q_\mathrm{U}^2 + U_\mathrm{U}^2}}{I_\mathrm{U}^2} - \dfrac{Q_\mathrm{U}Q_\perp + U_\perp U_\mathrm{U}}{I_\mathrm{U}\sqrt{Q_\mathrm{U}^2 + U_\mathrm{U}^2} } \right) \right.\\
    &\quad \quad \quad \left. + \sin{2\Psi_0 } \left( \dfrac{Q_\mathrm{U}Q_\times + U_\times U_\mathrm{U}}{I_\mathrm{U}\sqrt{Q_\mathrm{U}^2 + U_\mathrm{U}^2} } - \dfrac{I_\times \sqrt{Q_\mathrm{U}^2 + U_\mathrm{U}^2}}{I_\mathrm{U}^2} \right) \right] + \mathcal{o}(p_0^2) \\
    & \approx p_\mathrm{U} + p_0 \left[ \cos{2\Psi_0 } \left( \dfrac{I_\perp p_\mathrm{U}}{I_\mathrm{U}} - \dfrac{Q_\mathrm{U}Q_\perp + U_\perp U_\mathrm{U}}{p_\mathrm{U}I_\mathrm{U}^2 } \right) \right.\\
    &\quad \quad \quad \left. + \sin{2\Psi_0 } \left( \dfrac{Q_\mathrm{U}Q_\times + U_\times U_\mathrm{U}}{p_\mathrm{U}I_\mathrm{U}^2 } - \dfrac{I_\times p_\mathrm{U}}{I_\mathrm{U}} \right) \right] \textrm{ ,}
\end{split}
\end{equation}
where we used $p_\mathrm{U} \equiv \sqrt{Q_\mathrm{U}^2 + U_\mathrm{U}^2}/I_\mathrm{U}$ for the emergent polarization for unpolarized primary radiation. Due to symmetry, we may further assume $U_\mathrm{U} = U_\perp = U_\times = 0$. Denoting also $p_\perp \equiv Q_\perp/I_\perp$ and $p_\times \equiv Q_\times/I_\times$, one arrives at a simple decomposition of the total polarization degree:
\begin{equation}\label{any_Psi0}
    p = p_\mathrm{U} + p_0 \left[ \cos{2\Psi_0 } \dfrac{I_\perp}{I_\mathrm{U}} (p_\textrm{U} - p_\perp) + \sin{2\Psi_0 } \dfrac{I_\times}{I_\mathrm{U}}  (p_\times - p_\mathrm{U}) \right] \textrm{ .}
\end{equation}
The resulting total polarization direction $\Psi$ is given by the sign of $Q_\mathrm{tot}$, which is mainly driven by $Q_\mathrm{U}$ computed for unpolarized primary.

If we expect the primary polarization direction $\Psi_0$ to be only $0$ or $\pi/2$ (e.g. neglecting relativistic effects), we may use the notation for positive or negative polarization degree, respectively, and further reduce the expression:
\begin{equation}\label{small_p0}
    p = p_\mathrm{U} + p_0 \dfrac{I_\perp}{I_\mathrm{U}} (p_\mathrm{U}-p_\perp)\textrm{ .}
\end{equation}
Only two components are needed to estimate the resulting polarization degree for small $p_0$: the result for unpolarized primary and 100\% perpendicularly polarized primary in our case. These are to be precomputed either analytically or numerically for a particular reflection geometry.

\begin{figure*}
    \centering
    \includegraphics[width=2.\columnwidth]{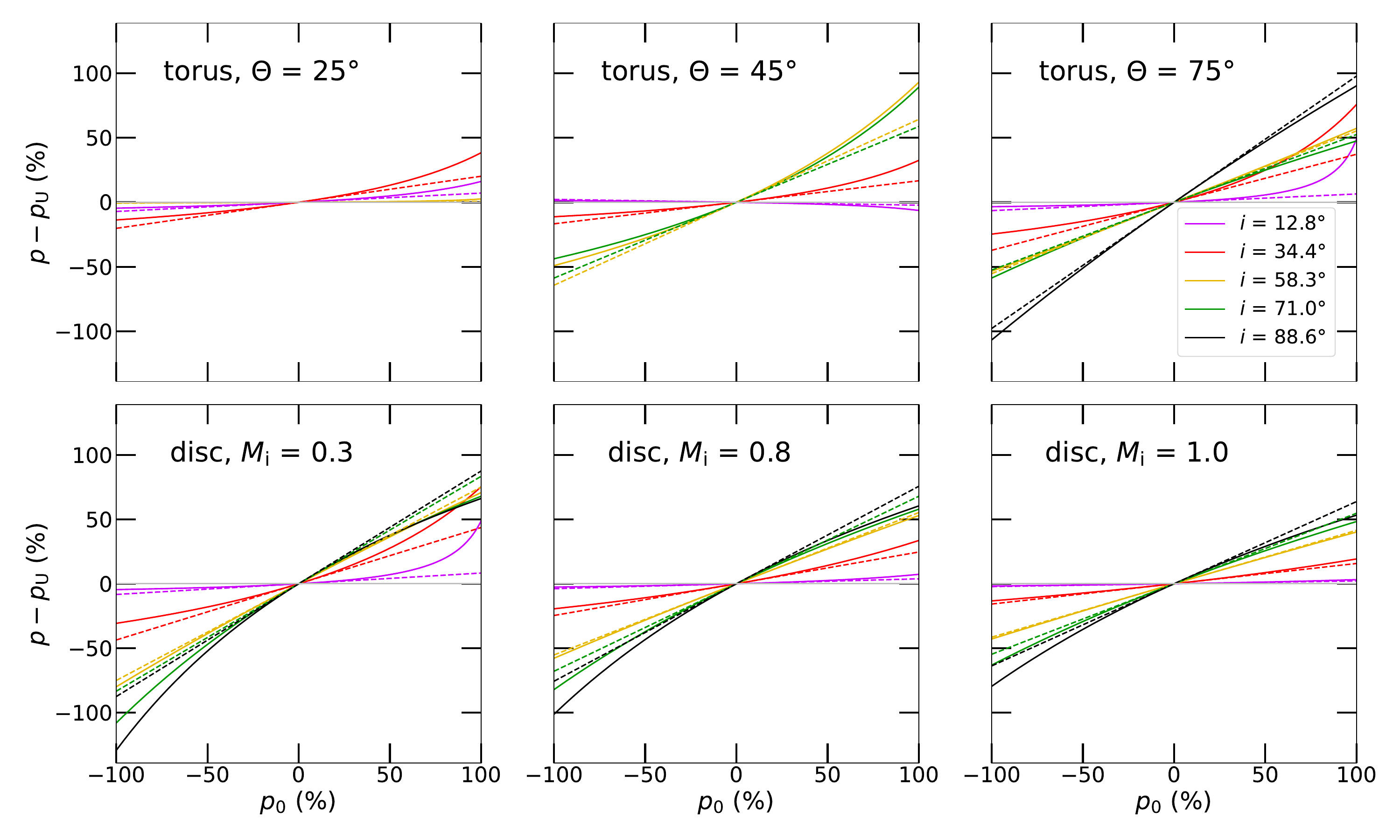}
    \caption{Comparison of the dependency of resulting polarization on incident polarization for the energy-independent single-scattering Chandrasekhar formulae integrated in the torus (top) and disc (bottom) geometries used in the {\tt xsstokes} model variants presented in this paper. From left to right we compare different values of $\Theta$ (the torus half-opening angle) and $M_\mathrm{i}$ (the disc illumination range) for the two reflection geometries, respectively. The results are shown for various observer's inclinations in the color code, which is another geometrical parameter that affects the net polarization response to incident polarization change. We plot the difference between the actual resulting polarization and the obtained polarization for unpolarized primary, $p-p_\mathrm{U}$, versus the polarization $p_0$ of the incident radiation, which is allowed to be either aligned with the rotation axis (positive) or orthogonal to it (negative). The solid lines are showing the exact result, the dashed lines represent linear approximation valid for small $p_0$, obtained from Equation (\ref{small_p0}). The results for high $\Theta$ of the torus approach the results for small $M_\mathrm{i}$ for the disc, as the scattering geometries become equivalent. The curves for high inclinations and small $\Theta$ for the torus are not shown, because the reflecting area is fully hidden, given the model geometrical setup.}
    \label{incident_results}
\end{figure*}

Figure \ref{incident_results} shows the full dependencies of $p - p_\mathrm{U}$ on $p_0$ computed with {\tt xsstokes} for the geometries assumed in this paper and using the energy-independent Chandrasekhar's single-scattering formulae for local reflection for simplicity. We also show the estimate for small $p_0$, using Equation (\ref{small_p0}). It holds that $(p_\mathrm{U} - p_\perp) \in [-2;2]$, but only for specific cases $\frac{I_\perp}{I_\mathrm{U}} \approx 1$, while usually $0.5 \lessapprox \frac{I_\perp}{I_\mathrm{U}} \lessapprox 2$. The overall factor $k = \frac{I_\perp}{I_\mathrm{U}} (p_\mathrm{U}-p_\perp)$ is $0 \lessapprox k \lessapprox 1$, which means that for small $p_0$, we may conclude that a smaller fraction of this primary polarization $p_0$ is typically added or subtracted (following its sign) to the result for $p_\mathrm{U}$ computed for unpolarized primary. If one releases the assumption of parallelly/perpendicularly oriented primary polarization direction with respect to the rotation axis, the deviation of the net result $p-p_\mathrm{U}$ by $k \cdot p_0$ is modulated by the $\cos{2\Psi_0 }$ and $\sin{2\Psi_0 }$ functions, according to Equation (\ref{any_Psi0}). Even when the parallelly/perpendicularly polarized primary assumption is in place, the difference $p-p_\mathrm{U}$ is not symmetric with respect to $p_0 = 0$ for larger $p_0$, i.e. the obtained full functions on Figure \ref{incident_results} are not odd.


\bsp	
\label{lastpage}
\end{document}